\documentclass[twocolumn,aip,reprint,superscriptaddress]{revtex4-2} 
\usepackage[T1]{fontenc}
\usepackage{amsmath,amsfonts,bm}

\usepackage{amssymb}
\usepackage{graphicx}  %epsfig,
\usepackage{color}
\usepackage{extarrows}
\usepackage{enumitem}
\usepackage{empheq}
\usepackage{tensor}
\usepackage[cal=rsfso]{mathalfa}
\usepackage{orcidlink}

\newcommand{\ud}{{\mathrm d}}

\newcommand{\ada}{{\rm ada}}
\newcommand{\hyb}{{\rm hyb}}
\newcommand{\gra}{{\rm gra}}

\newcommand{\w}{\omega}

\newcommand{\ti}{\tilde}

\newcommand{\B}{\mbox{\tiny B}}

\newcommand{\I}{\mbox{\tiny I}}

\newcommand{\tS}{\mbox{\tiny S}}
\newcommand{\T}{\mbox{\tiny T}}

\newcommand{\SB}{\mbox{\tiny SB}}
\newcommand{\dg}{\dagger}
\newcommand{\la}{\langle}
\newcommand{\ra}{\rangle}

\newcommand{\Sec}[1]{Sec.\,\ref{#1}}
\newcommand{\App}[1]{Appendix\,\ref{#1}}
\newcommand{\nl}{\nonumber \\}
\newcommand{\be}{\begin{equation}}
\newcommand{\ee}{\end{equation}}
\newcommand{\bsube}{\begin{subequations}}
\newcommand{\esube}{\end{subequations}}
\newcommand{\Eq}[1]{Eq.\,(\ref{#1})}
\newcommand{\Eqs}[1]{Eqs.\,(\ref{#1})}
\newcommand{\Fig}[1]{Fig.\,\ref{#1}}
\newcommand{\Figs}[1]{Figs.\,\ref{#1}}

\newcommand{\RN}[1]{%
  \textup{\uppercase\expandafter{\romannumeral#1}}%
}

\usepackage{siunitx}

\allowdisplaybreaks[1]

\definecolor{darkblue}{RGB}{0, 56, 102}

\usepackage{hyperref}  
\hypersetup{hidelinks,
	colorlinks=true,
	allcolors=darkblue,
	pdfstartview=Fit,
	breaklinks=true
}

\begin{document}

\title{Extended dissipaton theory with application to adatom-graphene composite
}
%%%

\author{Yu Su\,\orcidlink{0000-0003-4306-0161}}
\affiliation{
  Hefei National Research Center for Physical Sciences at the Microscale, University of Science and Technology of China, Hefei, Anhui 230026, China
}
\affiliation{
  Hefei National Laboratory,  University of Science and Technology of China, Hefei, Anhui 230088, China
}
\author{Yao Wang\,\orcidlink{0000-0003-2939-208X}}
\email{wy2010@ustc.edu.cn}
\affiliation{
  Hefei National Research Center for Physical Sciences at the Microscale, University of Science and Technology of China, Hefei, Anhui 230026, China
}
\affiliation{
  Hefei National Laboratory,  University of Science and Technology of China, Hefei, Anhui 230088, China
}
\author{Zi-Fan Zhu\,\orcidlink{0000-0002-3156-834X}}
\author{Yuan Kong\,\orcidlink{0000-0002-5637-2511}}
\affiliation{
  Hefei National Research Center for Physical Sciences at the Microscale, University of Science and Technology of China, Hefei, Anhui 230026, China
}
\affiliation{
  Hefei National Laboratory,  University of Science and Technology of China, Hefei, Anhui 230088, China
}
\author{Rui-Xue Xu\,\orcidlink{0000-0001-9931-201X}}
\email{rxxu@ustc.edu.cn}
\affiliation{
  Hefei National Research Center for Physical Sciences at the Microscale, University of Science and Technology of China, Hefei, Anhui 230026, China
}
\affiliation{
  Hefei National Laboratory,  University of Science and Technology of China, Hefei, Anhui 230088, China
}
\author{Xiao Zheng\,\orcidlink{0000-0002-9804-1833}}
\affiliation{
  Department of Chemistry, Fudan University, Shanghai 200433, China
}
\affiliation{
  Hefei National Laboratory,  University of Science and Technology of China,  Hefei, Anhui 230088, China
}
\author{YiJing Yan\,\orcidlink{0000-0002-4499-8154}}
\affiliation{
  Hefei National Research Center for Physical Sciences at the Microscale, University of Science and Technology of China, Hefei, Anhui 230026, China
}

\date{\today}

\begin{abstract}
  In this paper, we present the extended dissipaton theory, including the dissipaton-equation-of-motion formalism and the equivalent dissipaton-embedded quantum master equation. These are exact, non-Markovian, and non-perturbative theories, capable of handling not only linear but also quadratic environmental couplings. These scenarios are prevalent in a variety of strongly correlated electronic systems, including mesoscopic nanodevices and superconductors. As a demonstration, we apply the present theory to simulate the spectral functions of an adatom on a graphene substrate. We analyze the spectral peaks in the presence of the graphene substrate and compare them to those obtained in conventional metal environments. The adatom's spectral functions reveal intricate behaviors arising from the band structure of graphene.
\end{abstract}

\maketitle

\section{Introduction}\label{sec:intro}

Open quantum systems widely exist and play fundamental and crucial roles in various fields. Couplings between the primary functional site and its surroundings cause  particle, energy, and coherence exchange processes. Theoretical exploration on dynamic mechanism of open quantum systems helps design and control new devices. For example, recent years the electron transfer in graphene, particularly in the context of adatom-induced modifications on graphene's electronic structure and tunable magnetic properties, has emerged as a compelling direction.\cite{Sof12115405,Pik14115428,Ynd14245420,Shi19125158,Cao20045402,Gon16437} Understanding the electron transfer dynamics and the related correlation effects between the adatom and the graphene lattice is essential for advancing applications in spintronics, quantum computing, and sensor technologies, etc. However, the complexity of the underlying interactions, particularly in the presence of many-body effect, dissipation, and strong electron correlation, poses significant theoretical challenges. Besides, the Kondo effect, a many-body phenomenon arising from the interaction between a localized magnetic moment and the conduction electrons, has profound implications in nanoscale systems.\cite{Kon6437,Kon70183,Pus0145328,Swi03195318,Hew93}

The related theoretical investigations \cite{Mak14207401,Soh14125414,Cre15635019,Bul16270,Anv21155402} have been carried out by applying first-principle calculations combined with the numerical renormalization group method.\cite{Wil75773,Yos909403,Cos973003,Bul988365,Pet06245114,And08195216} Various other methods have been developed targeting on the equilibrium and dynamical properties of quantum impurities and other open quantum systems, such as the quantum Monte Carlo method, \cite{Gul11349,Har15085430} 
the multi-configuration time-dependent Hartree method (MCTDH), \cite{Mey9073,Wan031289,Zhe25052051}
the time-dependent renormalization group method, \cite{And05196801,Fri06144410} the time-dependent density matrix renormalization group method (TD-DMRG), \cite{Nis04613,Mer12075153,Ren22e1614} the time evolving density matrices using orthogonal polynomials algorithm, \cite{Pri10050404, Nus20155134} 
the quasi-adiabatic propagator path integral (QuAPI), \cite{Fey63118,Mak92435,Mak954600,Mak954611}
the time-evolving matrix product operator algorithm, \cite{Jor19240602,Ric22167403} the automated compression of environments method, \cite{Cyg22662} the inchworm quantum Monte Carlo method, \cite{Coh15266802, Che17054105, Cai202430} the quantum quasi-Monte Carlo algorithm, \cite{Ber21155104}  the auxiliary master equation approach, \cite{Sor19043303} and so on.

In the last decade, the dissipaton theory was developed as an exact formalism for open quantum dynamics,\cite{Yan14054105,Zha15024112,Wan22170901} which  utilizes the quasi-particle concept known as \emph{dissipaton} to describe the statistical properties and dynamical influences of the embedded thermal surrounding (bath). The system-plus-bath dynamics is expressed in terms of the interactions between the system and dissipatons. The dynamical equation established in this formalism, named as the dissipaton euqation of motion (DEOM), is an exact and nonperturbative approach for open quantum systems with Gaussian baths. The DEOM recovers the hierarchical equations of motion (HEOM) method \cite{Tan906676,Yan04216,Xu05041103,Jin08234703,Tan20020901} for the reduced system dynamics, but is more convenient and straightforward for simulating bath collective dynamics and polarizations.\cite{Zha15024112} The HEOM is the derivative equivalence of the Feymann--Vernon influence functional theory, which is applicable to the Gaussian environment linearly coupled to the system.\cite{Fey63118,Tan906676,Xu05041103} However, the dissipaton theory constructs the equation of motion just by applying the quasi-particle algebra to the total space von Neumann--Liouville equation.\cite{Yan14054105,Wan22170901}. The compact and convenient dissipaton algebra provides us potential to generalize the open system theories to more complicated scenarios. 
Compared with the other numerically exact time-dependent methods mentioned above, the dissipaton theory has the advantage in handling open quantum systems with strong non-Markovian bath couplings, achieving a balance between efficiency, accuracy and long-time stability through hierarchical expansions.

This paper presents the extended dissipaton theory for not only linear but also quadratic couplings.
The extended DEOM is constructed by generalizing the dissipaton algebra to the quadratic system-bath interactions.\cite{Su23024113}
The quadratic bath couplings are prevalent in a variety of strongly correlated electronic systems, including mesoscopic nanodevices\cite{Hew93,Phi12} and superconductors.\cite{Bal06373,Moc21186804}
The extended DEOM consitutes a powerful theoretical framework for modeling the nonequilibrium dynamics of complex quantum systems. Furthermore,  we also derive the extended dissipaton-embedded quantum master equation (DQME) for linear-plus-quadratic system-bath couplings. The DQME was originally developed only for  the case of linear system-bath couplings.\cite{Li24032620} In the extended DQME, instead of a hierarchical structure, all system–plus–dissipatons dynamics are incorporated into a single master equation. It provides further physical insights on the \emph{dissipatons}. That is, the dissipatons play the role of the statistical quasi-particles and characterize the primary interaction modes between the system and environment. 
For numerical demonstrations, we apply the extended dissipaton theory to study the electronic correlation effect in the adatom-graphene composite, with comparison to the metal substrate. We simulate the adatom spectral density to discuss the influence of the featured graphene band structure on the strongly correlated phenomena such as Kondo effect.

The remainder of this paper is organized as follows. In \Sec{thsecequ}, we present the general formalism of extended DEOM (details in \App{DEOM_derivation}) and the equivalent extended DQME (details in \App{appB}). In \Sec{thsecgra}, we build up the theoretical model of adatom-graphene composite (some details in \App{thappA}) and carry out numerical simulations. Finally, we summarize the paper in \Sec{thsum}. Throughout this paper, we set $\hbar=1$ and $\beta=1/(k_BT)$, with $k_B$ being the Boltzmann constant and $T$ being the temperature.

\section{Theory}\label{thsecequ}

In this section, we present the fermionic dissipaton theory. We start with the Hamiltonian settings, the dissipaton description of the bath influence, and the corresponding quasi-particle algebra. We then derive the extended dissipaton equation of motion (DEOM), containing both linear and quadratic system--bath interactions, together with the associated correlation function solver. Next, we give the extended dissipaton-embedded quantum master equation (DQME) as a quantum master equivalence of the DEOM, with both the system and dissipatons degrees of freedom involved. Finally we conclude the section with some comments.

\subsection{Total Hamiltonian}

Consider an electronic system ($H_{\tS}$) in contact with a fermionic bath ($h_{\B}$), where the total Hamiltonian reads
\be \label{total}
H_{\T}=H_{\tS}+H^{\I}_{\SB}+H^{\I\I}_{\SB}+h_{\B}.
\ee
In \Eq{total}, while $H_{\tS}$ is arbitrary, the bath Hamiltonian $h_{\B}$ is modeled as non-interacting electrons,
\begin{align}\label{hB}
  h_{\B} = \sum_{ks} \epsilon_{ks}\hat d^+_{ks}\hat d_{ks}^-,
\end{align}
where $k$ and $s=\uparrow,\downarrow$ label a single--electron spin--orbital state of the bath.
The system and bath couple with each other via linear and quadratic interactions, reading 
\begin{align}\label{SB1}
  H^{\I}_{\SB} = \sum_{us}(\hat a^+_{us}\hat F_{us}^-+\hat F^+_{us}\hat a_{us}^-) \equiv \sum_{\sigma us}\hat a^{\bar\sigma}_{us}\tilde{F}^\sigma_{us},
\end{align} 
and
\begin{align}\label{SB2}
  H^{\I\I}_{\SB} = \frac{1}{2}\sum_{\sigma us}\sum_{\sigma' u's'}\hat q^{\bar\sigma\bar\sigma'}_{us,u's'}\tilde{F}^{\sigma}_{us}\tilde{F}^{\sigma'}_{u's'},
\end{align}
respectively.
In \Eq{SB1}, $\hat a^+_{us}\equiv(\hat a^-_{us})^\dagger$, where $\{\hat a^\sigma_{us}\}$ are system creation ($\sigma=+$) and annihilation ($\sigma=-$) operators and $u$ denotes the specific electronic orbital state.
The hybridizing bath operators read 
\begin{align}\label{FB}
  \hat F_{us}^+\equiv\sum_k t_{kus}\hat d_{ks}^+\equiv(\hat F^-_{us})^\dagger.
\end{align}
Besides, \Eq{SB1} defines 
\be 
\tilde{F}^\sigma_{us}\equiv-\sigma\hat F^\sigma_{us}\equiv\bar\sigma\hat F^\sigma_{us}
\ee
for notational convenience.
Equation (\ref{SB2}) involves also the system subspace operators $\{\hat q^{\sigma\sigma'}_{us,u's'}\}$. %which are generally quadratic with respect to $\{\hat a_{us}^\sigma\}$. It is closely related to the form of two--particle interactions in many--electron systems. 
Without loss of generality, we set $\{\hat q^{\sigma\sigma'}_{us,u's'}\}$ to be antisymmetric,\cite{Su23024113} 
\begin{align}\label{qanti}
  \hat q^{\sigma\sigma'}_{us,u's'}=-\hat q^{\sigma'\sigma}_{u's',us}.
\end{align}
%\be 
%\hat q^{\sigma\sigma'}_{us,u's'}=(q^{\bar \sigma\bar \sigma'}_{us,u's'})^{\dg}
%\ee
% The antisymmetric condition can be generally achieved by noticing that
% \begin{align}
%   \{ \hat F^+_{us}, \hat F^-_{u's'} \} = \delta_{ss'}\sum_k t_{kus}t^*_{ku's}
% \end{align}
% is a c-number. 

For the environment and its coupling given by \Eqs{hB} and (\ref{FB}), its influence is completely described by the hybridizing bath spectral density functions,
\begin{align}
  \Gamma_{uvs}(\omega) \equiv \Gamma^-_{uvs}(\omega) = \pi\sum_{k}t_{kus}^*t_{kvs}\delta(\omega-\epsilon_{ks}),
\end{align}
which can be equivalently expressed via 
\begin{align}
  \Gamma^\sigma_{uvs}(\omega) \equiv \frac{1}{2}\int_{-\infty}^\infty\!\!\ud t\, e^{-i\sigma\omega t}\la\{ \hat F^\sigma_{us}(t),\hat F^{\bar\sigma}_{vs}(0) \}\ra_{\B},
\end{align}
with $\Gamma^\sigma_{vus}(\omega)=[\Gamma^\sigma_{uvs}(\omega)]^*=\Gamma^{\bar\sigma}_{uvs}(\omega)$. Here, we follow the bare-bath thermodynamic prescription: $\hat F^\sigma_{us}(t)\equiv e^{ih_{\B}t}\hat F^\sigma_{us}e^{-ih_{\B}t}$ and $\la\hat O\ra_{\B}\equiv {\rm tr}_{\B}(\hat Oe^{-\beta h_{\B}})/{\rm tr_{\B}}e^{-\beta h_{\B}}$. We then have 
\begin{align}\label{FDT1}
  \la\hat F^\sigma_{us}(t)\hat F^{\bar\sigma}_{vs'}(0)\ra_{\B} = \frac{\delta_{ss'}}{\pi}\int_{-\infty}^\infty\!\!\ud\omega\, e^{\sigma i\omega t}\frac{\Gamma^\sigma_{uvs}(\omega)}{1+e^{\sigma\beta\omega}},
\end{align}
the fermionic fluctuation--dissipation theorem.
%\cite{Yan14054105} 

\subsection{Dissipaton description}

To proceed, we expand the bath correlation function as sum of exponential functions, 
%\bsube
\begin{align}\label{Fc}
  \la\hat F^\sigma_{us}(t)\hat F^{\bar\sigma}_{vs}(0)\ra_{\B} = \sum_{\kappa=1}^K\eta^\sigma_{\kappa uvs}e^{-\gamma^\sigma_{\kappa us}t},
\end{align}
whose time reversal reads
\begin{align}\label{Fcc}
  \la\hat F^{\bar\sigma}_{vs}(0)\hat F^\sigma_{us}(t)\ra_{\B} = \sum_{\kappa=1}^K\eta^{\bar\sigma*}_{\kappa uvs}e^{-\gamma^\sigma_{\kappa us}t},
\end{align}
%\esube
with the property $\gamma^\sigma_{\kappa us}=(\gamma^{\bar\sigma}_{\kappa us})^*$ being required in the above exponential decomposition.%\cite{Yan16110306}
The construction of the dissipaton formalism starts from the exponential series expansion as \Eq{Fc}.
It is 
evaluated from \Eq{FDT1}
via the Cauchy’s
residue theorem in contour integration.
The integration via residues depends on not only the concrete form of the spectral density, but also the fractional decomposition of the fermionic function.
For the latter,
traditionally, people adopt the Mittag--Leffler decomposition, specifically named also as the Matsubara expansion for the distribution function.
This traditional scheme is however very slow in convergence.
By far, one of the most efficient expansions of \Eq{Fc} is the 
time-domain Prony
fitting decomposition ($t$-PFD)\cite{Che22221102}
which is applied to arbitrary correlation functions in the time domain. 

Within the HEOM method, the exponential decomposition [\Eq{Fc}] is a scheme to making the equations closed when performing derivative to the Feynmann--Vernon influence functional.\cite{Tan906676} In the dissipaton theory, we treat each exponential function $\eta_{\kappa uvs}^\sigma e^{-\gamma_{\kappa us}^\sigma t}$ as a generalized quasi-particle with an effective complex eigenfrequency $-i\gamma^{\sigma}_{\kappa us}$. Here, the real and imaginary parts of $\gamma^{\sigma}_{\kappa us}$ stand for the oscillation and dissipation motions, respectively. To proceed, we decompose 
\begin{align}
  \tilde{F}^\sigma_{us} = \sum_{\kappa=1}^K\hat f^\sigma_{\kappa us},
\end{align}
with 
\begin{subequations}\label{fcf}
  \begin{align}
    &\la\hat f^\sigma_{\kappa us}(t)\hat f^{\sigma'}_{\kappa' vs'}(0)\ra_{\B}\! =\la\hat f^\sigma_{\kappa us}\hat f^{\sigma'}_{\kappa' vs'}\ra_{\B}^{>}\,e^{-\gamma^\sigma_{\kappa us}t},\\
    &\la\hat f^{\sigma'}_{\kappa'vs'}(0)\hat f^\sigma_{\kappa us}(t)\ra_{\B}\! =\la\hat f^{\sigma'}_{\kappa'vs'}\hat f^\sigma_{\kappa us}\ra_{\B}^{<}\,e^{-\gamma^{\sigma}_{\kappa us}t},
  \end{align}
\end{subequations}
where
\[
\la\hat f^\sigma_{\kappa us}\hat f^{\sigma'}_{\kappa' vs'}\ra_{\B}^{>}\equiv \la\hat f^\sigma_{\kappa us}(0+)\hat f^{\sigma'}_{\kappa' vs'}\ra_{\B}=\!-
    \delta_{\sigma\bar\sigma'}\delta_{\kappa\kappa'}\delta_{ss'}\eta^\sigma_{\kappa uvs}
\]
and
\[
\la\hat f^{\sigma'}_{\kappa'vs'}\hat f^\sigma_{\kappa us}\ra_{\B}^{<}\equiv\la\hat f^{\sigma'}_{\kappa'vs'}\hat f^\sigma_{\kappa us}(0+)\ra_{\B}=-\delta_{\sigma\bar\sigma'}\delta_{\kappa\kappa'}\delta_{ss'}\eta^{\bar\sigma*}_{\kappa uvs}.
\]
Here, $\{\hat f^\sigma_{\kappa us}\}$ are denoted as the dissipaton operators, providing a statistical quasi--particle picture to account for environmental influences. It is evident that \Eq{fcf} can reproduce both \Eqs{Fc} and (\ref{Fcc}). 
For simplicity, we adopt the index abbreviations,
\begin{align}\label{jj}
  j\equiv (\sigma\kappa us)
\end{align}
and $\bar j\equiv(\bar\sigma\kappa us)$, 
leading to $\hat f_j\equiv \hat f^\sigma_{\kappa us}$ and so on. Then we can recast \Eqs{SB1} and (\ref{SB2}) into 
\begin{align}\label{SB1f}
  H_{\SB}^{\I} = \sum_j\hat a_{\bar j}\hat f_j
\end{align}
and 
\begin{align}\label{SB2f}
  H_{\SB}^{\I\I} = \frac{1}{2}\sum_{jj'}\hat q_{\bar j\bar j'}\hat f_j\hat f_{j'},
\end{align}
respectively. Here, we define $\hat a_j\equiv\hat a^\sigma_{us}$ and $\hat q_{j j'}\equiv \hat q^{\sigma\sigma'}_{us,u's'}$.

\subsection{The extended DEOM}

Dissipaton operators $\{\hat f_{j}\}$, together with the total density operator $\rho_{\T}(t)$, are used to define the dissipaton density operators (DDOs),  dynamical variables in the DEOM. The DDOs are defined as
\begin{align}
  \rho_{\bf j}^{(n)}(t)\equiv \rho_{j_1\cdots j_n}^{(n)}(t)\equiv {\rm tr}_{\B}[(\hat f_{j_n}\cdots \hat f_{j_1})^\circ\rho_{\T}(t)].
\end{align}
The notation, $(\cdots)^\circ$, denotes the \textit{irreducible} dissipaton product notation, with $(\hat f_j\hat f_{j'})^\circ=-(\hat f_{j'}\hat f_j)^\circ$ for fermionic dissipatons.
The subscript ${\bf j}\equiv j_1\cdots j_n$ describes a fermionic dissipaton configuration, where each $j_r$ specifies a set of values defined in \Eq{jj}.
 The DDOs for fermionic coupled
environment resemble a Slater determinant, having the
occupation number of $0$ or $1$ only, due to the antisymmetric fermion permutation relation.
The reduced system density operator is just $\rho^{(0)}_{}(t)={\rm tr}_{\B}[\rho_{\T}(t)]\equiv \rho_{\tS}(t)$. 

The extended DEOM, with both linear and quadratic system-bath interactions, reads 
\begin{widetext}
  \begin{align}\label{DEOM_wy0828}
    \dot{\rho}_{\bf j}^{(n)} &= -\bigg(i\mathcal{L}_{\tS}^{\rm eff}+\sum_{r=1}^n\gamma_{j_r}\bigg)\rho_{\bf j}^{(n)}+i\sum_{r=1}^n\sum_{j}(-)^{n-r}\mathcal{B}_{j_rj}\rho_{{\bf j}_r^-j}^{(n)} -i\sum_j\tensor{\mathcal{A}}{_{\bar j}}\rho_{{\bf j}j}^{(n+1)}-i\sum_{r=1}^n(-)^{n-r}\tensor{\mathcal{C}}{_{j_r}}\rho_{{\bf j}_r^-}^{(n-1)} \nl
    &\quad -\frac{i}{2}\sum_{jj'}\tensor{\mathcal{A}}{_{\bar j\bar j'}}\rho_{{\bf j}j'j}^{(n+2)}-i\sum_{r>r'}(-)^{r-r'}\tensor{\mathcal{C}}{_{j_rj_{r'}}}\rho_{{\bf j}_{rr'}^{--}}^{(n-2)}.
  \end{align}
\end{widetext}
Here, the superoperators on $\{\rho^{(n)}\}$ parts are defined as
\begin{align}\label{eq19}
\mathcal{L}^{\rm eff}_{\tS}
\hat O &\equiv [H_{\tS}^{\rm eff},\hat O]\equiv [H_{\tS}+\la H_{\SB}^{\I\I}\ra_{\B},\hat O], 
\end{align}
and
\begin{align} 
\!\!\!\!\!\mathcal{B}^{\sigma,\sigma'}_{\kappa us,u's'}\hat O &\equiv \sum_v \Big(\eta^{\sigma}_{\kappa uvs}\hat{q}^{\sigma\bar\sigma'}_{vs,u's'}\hat O+\eta^{\bar\sigma*}_{\kappa uvs}\hat O\hat{q}^{\sigma\bar\sigma'}_{vs,u's'}\Big),
\end{align}
with [cf.\,\Eq{SB2}]
\be 
\la H_{\SB}^{\I\I}\ra_{\B}=\frac{1}{2}\sum_{\sigma s}\sum_{uv}\la\tilde F^\sigma_{us}\tilde F^{\bar\sigma}_{vs}\ra_{\B}\hat q^{\bar\sigma \sigma}_{us,vs},
\ee
via \Eq{FDT} at $t=0$ for $\la\tilde F^\sigma_{us}\tilde F^{\bar\sigma}_{vs}\ra_{\B}$.
In \Eq{DEOM_wy0828}, actions on the $\{\rho^{(n\pm 1)}\}$ parts include
\begin{subequations}\label{AS}
  \begin{align}
    \tensor*{\mathcal{A}}{_{us}^\sigma}\rho^{(n\pm 1)} &\equiv \hat a^\sigma_{us}\rho^{(n\pm 1)} - (-)^{n}\rho^{(n\pm 1)}\hat a_{us}^\sigma,\\ 
    \tensor*{\mathcal{C}}{_{\kappa us}^\sigma}\rho^{(n\pm 1)} &\equiv \sum_v\Big[\eta^\sigma_{\kappa uvs}\hat a^\sigma_{vs}\rho^{(n\pm 1)}\nl
    &\qquad\quad + (-)^{n}\eta^{\bar\sigma*}_{\kappa uvs}\rho^{(n\pm 1)}\hat a^\sigma_{vs}\Big],
  \end{align}
\end{subequations}
whereas those on the $\{\rho^{(n\pm 2)}\}$ parts include
\begin{subequations}
  \begin{align}
    \tensor*{\mathcal A}{^{\sigma,\sigma'}_{us,vs'}}\hat O &\equiv [\hat{q}^{\sigma\sigma'}_{us,vs'},\hat O],\\
    \label{eq23b}\tensor*{\mathcal C}{_{\kappa us,\kappa'vs'}^{\sigma,\sigma'}}\hat O &\equiv \sum_{u'v'}\Big(\eta^{\sigma}_{\kappa uu' s}\eta^{\sigma'}_{\kappa'vv's'}\hat{q}^{\sigma\sigma'}_{u's,v's'}\hat O\nl
    &\quad\quad\quad - \eta^{\bar\sigma\ast}_{\kappa uu's}\eta^{\bar\sigma'*}_{\kappa'vv's'}\hat O\hat{q}^{\sigma\sigma'}_{u's,v's'}\Big).
  \end{align}
\label{eq23}
\end{subequations}
The detailed derivations of \Eq{DEOM_wy0828} with \Eqs{eq19}--(\ref{eq23}) are given in \App{DEOM_derivation}.
The DEOM is composed of a set of  hierarchically coupled linear differential equations of DDOs. Formally DEOM consists of an infinite hierarchy and needs to be truncated in practice.

\subsection{Correlation function solver}\label{subsection:2.4}

By \Eq{DEOM_wy0828}, we can obtain the transient dynamics and steady state of DDOs.
Furthermore, the dissipaton theory also serves as a solver for evaluating correlation functions, 
\begin{align}\label{css}
    \la \hat a_s^-(t) \hat a_s^+(0) \ra \equiv {\rm Tr}\big[ \hat a_s^-e^{-i H_{\T}t}(\hat a_s^+\rho_{\T}^{\rm ss})e^{i H_{\T}t} \big],
\end{align}
where $\rho_{\T}^{\rm ss}$ is the steady state of the total system, $H_{\T}$. The algorithm for solving \Eq{css} is as follows. (i) Obtain the steady state in the dissipaton representation, $\{\rho_{\bf j}^{(n);\rm ss} \equiv {\rm tr}_{\B}[(\hat f_{j_n}\cdots \hat f_{j_1})^\circ\rho_{\T}^{\rm ss}]\}$, by propagating the extended DEOM or solving $\dot\rho_{\bf j}^{(n);\rm ss} = 0$ directly. (ii) Define new dynamical variables as 
\begin{align}\label{varrho}
    \varrho_{\bf j}^{(n)}(t) \equiv {\rm tr}_{\B}[(\hat f_{j_n}\cdots \hat f_{j_1})^\circ e^{-iH_{\T}t}( \hat a_s^+\rho_{\T}^{\rm ss} )e^{iH_{\T}t}].
\end{align}
We then have its initial state being 
\begin{align}
    \varrho_{\bf j}^{(n)}(0) = (-)^n \hat a_s^+ {\rm tr}_{\B}[(\hat f_{j_n}\cdots \hat f_{j_1})^\circ \rho_{\T}^{\rm ss} ] = (-)^n \hat a_s^+\rho_{\bf j}^{(n);\rm ss}.
\end{align}
The time evolution of \Eq{varrho} can be derived as 
\begin{align}
    \dot{\varrho}_{\bf j}^{(n)} &= -\bigg(i\mathcal{L}_{\tS}^{\rm eff}+\sum_{r=1}^n\gamma_{j_r}\bigg)\varrho_{\bf j}^{(n)}+i\sum_{r=1}^n\sum_{j}(-)^{n-r}\mathcal{B}_{j_rj}\varrho_{{\bf j}_r^-j}^{(n)}\nl 
    &\quad -i\sum_j\tensor{\widetilde{\mathcal{A}}}{_{\bar j}}\varrho_{{\bf j}j}^{(n+1)}-i\sum_{r=1}^n(-)^{n-r}\tensor{\widetilde{\mathcal{C}}}{_{j_r}}\varrho_{{\bf j}_r^-}^{(n-1)} \nl
    &\quad -\frac{i}{2}\sum_{jj'}\tensor{\mathcal{A}}{_{\bar j\bar j'}}\varrho_{{\bf j}j'j}^{(n+2)}-i\sum_{r>r'}(-)^{r-r'}\tensor{\mathcal{C}}{_{j_rj_{r'}}}\varrho_{{\bf j}_{rr'}^{--}}^{(n-2)},
\end{align}
where 
\begin{subequations}
  \begin{align}
    \tensor*{\widetilde{\mathcal{A}}}{_{us}^\sigma}\varrho^{(n\pm 1)} &\equiv \hat a^\sigma_{us}\varrho^{(n\pm 1)} + (-)^{n}\varrho^{(n\pm 1)}\hat a_{us}^\sigma,\\ 
    \tensor*{\widetilde{\mathcal{C}}}{_{\kappa us}^\sigma}\varrho^{(n\pm 1)} &\equiv \sum_v\Big[\eta^\sigma_{\kappa uvs}\hat a^\sigma_{vs}\varrho^{(n\pm 1)} \nl
    &\quad\qquad - (-)^{n}\eta^{\bar\sigma*}_{\kappa uvs}\varrho^{(n\pm 1)}\hat a^\sigma_{vs}\Big],
  \end{align}
\end{subequations}
which differs from \Eqs{DEOM_wy0828} and (\ref{AS}) with an extra minus sign due to the odd parity of $\hat a_s^+\rho_{\T}^{\rm ss}$. 
The superoperators are same as \Eq{eq23} for $\varrho^{(n\pm 2)}$.
(iii) Calculate the correlation function by 
\begin{align}
    \la \hat a_s^-(t) \hat a_s^+(0) \ra = {\rm tr}_{\tS}[\hat a_s^-\varrho_{}^{(0)}(t)]. 
\end{align} 
For the counterpart $\la\hat a_s^+(0)\hat a_s^-(t)\ra$, we adopt a similar procedure just by redefining $\varrho_{\bf j}^{(n)}(t) \equiv {\rm tr}_{\B}[(\hat f_{j_n}\cdots \hat f_{j_1})^\circ e^{-iH_{\T}t}( \rho_{\T}^{\rm ss}\hat a_s^+ )e^{iH_{\T}t}]$. We thus establish the correlation function solver based on the dissipaton theory.

\subsection{The extended DQME}

In this subsection, we present the extended dissipaton-embedded quantum master equation (DQME), where the system-bath interactions involve both the linear and quadratic terms. The DQME formalism represents the collective motion concerning the system as well as dissipatons in the form of quantum master equation. For clarity, we consider the case that $\la\hat F^\sigma_{us}(t)\hat F^{\bar\sigma}_{vs}(0)\ra_{\B} = \delta_{uv}\la\hat F^\sigma_{us}(t)\hat F^{\bar\sigma}_{us}(0)\ra_{\B} = \delta_{uv}\sum_{\kappa}\eta^\sigma_{\kappa us}e^{-\gamma^\sigma_{\kappa us}t}$. Then, the extended fermionic DQME reads 
\begin{widetext}
\begin{align}\label{eDQME}
  \dot{\tilde\rho} &= -\Big(iH_{\tS}^{\rm eff}+ \sum_k \gamma_k^-\hat N_k\Big)\tilde\rho -i \sum_k\Big( \zeta_k^- [\hat a_k^+,\hat b_k^-\tilde\rho ] + \xi_k^+\hat a_k^+\tilde\rho\hat b_k^-  - \xi_k^-\hat a_k^-\hat b_k^+\tilde\rho \Big)\nl  
  &\quad\,-i\sum_{kk'}\bigg\{ \frac{1}{2}\zeta_k^-\zeta_{k'}^-[\hat q_{kk'}^{--},\tilde\rho\hat b_k^+\hat b_{k'}^+] + \frac{1}{2}\zeta_{k}^-\zeta_{k'}^+[\hat q^{+-}_{kk'},\hat b_k^-\tilde\rho\hat b_{k'}^+]  \nl 
  &\quad\, +\zeta_k^+\xi_{k'}^-( \hat q_{kk'}^{--}\hat b_{k'}^+\tilde\rho\hat b_k^+ -\hat q_{kk'}^{+-}\hat b_k^-\hat b_{k'}^+\tilde\rho) +\zeta_k^-\xi_{k'}^+( \hat q_{kk'}^{++}\hat b_k^-\tilde\rho\hat b_{k'}^--\hat q_{kk'}^{-+}\tilde\rho\hat b^-_{k'}\hat b^+_{k} )\nl
  &\quad\, + \frac{1}{2}\xi_k^+\xi_{k'}^+\hat q_{kk'}^{++}\tilde\rho\hat b_{k}^-\hat b_{k'}^-  + \frac{1}{2}\xi_k^-\xi_{k'}^-\hat q^{--}_{kk'}\hat b_k^+\hat b_{k'}^+\tilde\rho \nl 
  &\quad\,+\xi_{k}^+\xi_{k’}^-\hat q^{+-}_{kk'}\hat b_{k'}^+\tilde\rho\hat b_{k}^- \bigg\}
  + {\rm g.h.c.},
\end{align}
\end{widetext}
where the generalized Hermitian conjugate (g.h.c), denoted via $\ddag$, is defined as 
\begin{equation}\label{ghc}
\begin{split}
&(\gamma_{k}^{\sigma})^{\ddag}=\gamma_{k}^{\bar \sigma},\ \ 
(\hat a_{k}^{\sigma})^{\ddag}=\hat a_{k}^{\bar \sigma},\ \ 
(\hat b_{k}^{\sigma})^{\ddag}=\hat b_{k}^{\bar \sigma},
\\
&(\hat q_{kk'}^{\sigma\sigma'})^{\ddag}=\hat q_{k'k}^{\bar\sigma'\bar\sigma},\ \ (\ti\rho)^{\ddag}=\ti\rho.
\end{split}  
\nonumber
\end{equation}
The involved coefficients are $\zeta_k^\sigma\equiv(\eta_k^\sigma\eta_k^{\bar\sigma*})^{1/4}=(\zeta_k^{\bar\sigma})^{\ddag}$ and $\xi_k^\sigma \equiv \eta_k^\sigma / \zeta_k^\sigma$ with $(\xi_k^\sigma)^{\ddag}=(\xi_k^\sigma)^{\ast}$. The g.h.c.\,here is equivalent to the definition of Hermitian superoperator $\mathcal O^\dagger = \mathcal O$, in relation to $\mathcal O^\dagger\rho \equiv (\mathcal O\rho^\dagger)^\dagger$ for arbitrary $\rho$. In \Eq{eDQME}, $\{\hat b_k^\sigma\}$ are the fermionic creation ($\sigma=+$) and annihilation ($\sigma=-$) operators for the dissipaton embedded degrees of freedom, satisfying
\begin{align}
    \{\hat b_k^{-}, \hat b_{k'}^-\} = \{\hat b_k^{+}, \hat b_{k'}^+\} = 0\quad\text{and}\quad \{\hat b_k^{-}, \hat b_{k'}^+\} = \delta_{kk'},
\nonumber
\end{align}
and the number operators are defined as $\hat N_k \equiv \hat b_k^+\hat b_k^-$. Here, the index $k$ represents the collection $(\kappa us)$, i.e., $j = (\sigma\kappa us) = (\sigma, k)$.  In the absence of quadratic system--bath coupling, the extended DQME \Eq{eDQME} reduces to the form of Eq.\,(15) of Ref.\citenum{Li24032620}. The detailed derivations are in \App{appB}.
The extended DQME, instead of a hierarchical structure, \Eq{DEOM_wy0828}, incorporates all system–plus–dissipaton degrees of freedom into a single master equation, which is more versatile to accommodate quantum algorithms.\cite{Li24032620}

%and many-body ansatz.\cite{Zha23080901}

\subsection{Some comments}
To conclude this section, we would like to comment on the constrictions of the present formalism.
Although the dissipaton theory presented above have achieved breakthroughs in describing environmental structures from linear to nonlinear couplings, the environment itself is still required to consist of non-interacting electrons [cf.\,\Eq{hB}]. 
Notably, it has been shown the isomorphism between non-interacting electrons and multi-state systems\cite{Liu17024110}. 
Similarly, in the corresponding bosonic version of the theory, we also require the environment to be harmonic, i.e., composed of free bosons.\cite{Xu18114103,Che23074102}

While this bath assumption of free particles already encompasses a broad class of open quantum systems, the environment of interacting (quasi-)particles exist in reality.
The dissipaton theory encounters challenges when dealing with such environments.
For instance, in condensed-phase chemical reactions, if the anharmonic effects of solvent vibrations are non-negligible, they lead to interactions between vibrational excitations. In such cases, the solvent environment would not be transformed to a free-particle system.\cite{Got152722}
Another example is that, although photons in quantum electrodynamics do not interact with each other and can be treated as free particles, gluons in quantum chromodynamics exhibit strong interactions and cannot be considered free particles, despite often being approximated as such in the literature.\cite{Yao212130010} 
In fermionic scenarios, the Fermi-Hubbard bath introduces electron-electron interactions, \cite{Wei25_arxiv_2501_05562} which goes beyond the non-interacting bath assumption and have to be treated using mean-field approximations.
These scenarios pose new challenges for quantum dissipation theory.
When the bath cannot be described by simple treatments of harmonic bath or non-interacting fermionic bath, one may deal with bath degrees of freedom explicitly.  
In such cases, some precise approaches or practical approximate methods have been developed to tackle the problem.\cite{Yan19074106,Liu17102319,He22e1619,Wu24644,He245452,Wei25_arxiv_2501_05562}

\section{Numerical demonstrations}\label{thsecgra}

In this section,
we apply the dissipaton theory to study the 
adatom-graphene composite by simulating the adatom spectral density.
Graphene is characterized by its two-dimensional hexagonal lattice of carbon atoms and delocalized $\pi$-bonding across the entire structure.
This unique structure results in exceptional mechanical strength, high electron mobility, and excellent thermal capacity and conductivity.
It has garnered widespread interest due to its diverse properties, functionalities, and potential real-world applications. 
\cite{Cas09109}
Among its many intriguing characteristics,
the tuning effects induced by adsorbed atoms (adatoms) provide immense potential for the design of graphene-based electronic devices.
 Since the extended DEOM and the extended DQME are equivalent to each other, one may simulate the open system using either approach.

\subsection{Model of adatom-graphene composite}\label{model}

%\subsection{Graphene Hamiltonian}

Consider the model of adatom-graphene composite, whose Hamiltonian reads \cite{Ynd14245420}
\begin{align}
H&=H_{\ada}+H_{\hyb}+H_{\gra}.
\end{align}
Here, the adatom Hamiltonian reads
\be \label{hs}
H_{\ada}=\sum_{s=\uparrow,\downarrow}\epsilon_{s}\hat n_{s}+U\hat n_{\uparrow}\hat n_{\downarrow}
\ee
with $\hat n_{s}=\hat a_{s}^{\dg}\hat a_{s}$, where $\hat a_{s}^{\dg}$ and $\hat a_{s}$ are the electronic creation and annihilation operators in the adatom with $s$-spin.
$H_{\gra}$ and  $H_{\hyb}$ are the graphene and the adatom-graphene  hybridization Hamiltonians, respectively, which will be elaborated later.

\begin{figure}[t]  
\centering\includegraphics[width=0.7\columnwidth]{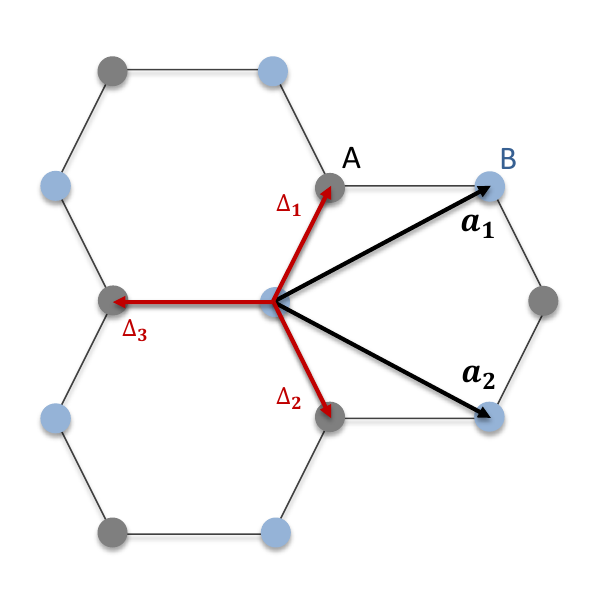} 
\caption{Several cells of the infinitely extended lattice structure of graphene, with two sublattice $A$ and $B$. Here, ${\bm a}_1$ and ${\bm a}_2$ are two basis vectors. The three $\Delta$-vectors are used to compute $\theta_{\bm k}$ in \Eq{h_graph_2}.}      
\label{fig1}   
\end{figure}

\begin{figure}[t]  
\centering\includegraphics[width=0.85\columnwidth]{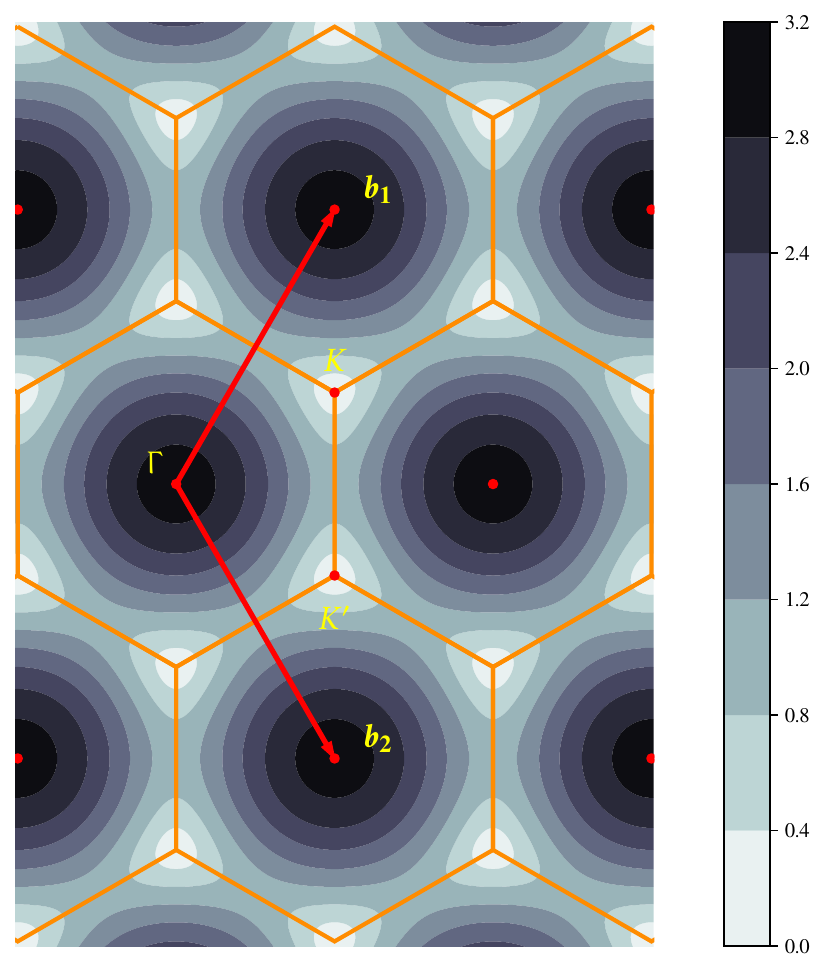} 
\caption{Reciprocal lattice structure of graphene and the value of $\theta_{\bm k}$. Here, ${\bm b}_1$ and ${\bm b}_2$ are two reciprocal lattice vectors. The Dirac cones are located at  $K$ and $K'$ points, where $\theta_{\bm k}=0$.}       
\label{fig2}   
\end{figure}

For the graphene part, under the tight-binding approximation, only the transitions between nearest neighbors are considered for the electrons.
Its Hamiltonian adopts \cite{Cas09109}
\begin{align}\label{graph}
H_{\gra}=-g\sum_{\la m,n \ra, s}\Big(\hat\alpha_{m s}^{\dg}\hat\beta_{n s}+\hat\beta_{n s}^{\dg}\hat\alpha_{m s}\Big),
\end{align}
where $g$ is the transfer coupling constant, and  $(\hat\alpha^{\dg}_{m s}, \hat\alpha_{m s})$ and $(\hat\beta^{\dg}_{ms}, \hat\beta_{m s})$ are the creation and annihilation operators for electrons with $s$-spin at $m$-site of $A$ and $B$ sublattices, respectively (cf.\,\Fig{fig1}).
Graphene is a single layer of carbon atoms arranged in a two-dimensional honeycomb lattice .
The lattice can be described using two basis vectors (cf.\,\Fig{fig1}):
\begin{align*}
\bm {a}_1=\frac{a}{2}(\sqrt{3},1),\ \ \bm {a}_2=\frac{a}{2}(\sqrt{3},-1).
\end{align*}
Here, 
$a=|\bm {a}_1|=|\bm {a}_2|$ is the lattice constant, which is approximately 2.46$\si{\angstrom}$.
The reciprocal lattice of graphene is also a honeycomb structure, with reciprocal lattice vectors (cf.\,\Fig{fig2}):
\begin{align*}
\bm {b}_1=\frac{2\pi}{a}(\frac{1}{\sqrt{3}},1),\ \ \bm {b}_2=\frac{2\pi}{a}(\frac{1}{\sqrt{3}},-1),
\end{align*}
such that $\bm {a}_i\cdot \bm {b}_j=2\pi\delta_{ij}$.

By setting
\begin{align}
\begin{bmatrix}
\hat \alpha_{ns}\\
\hat \beta_{ns}\\
\end{bmatrix}
=\frac{1}{\sqrt{N}}\sum_{{\bm k}\in {\rm BZ}}e^{-i{\bm k}\cdot {\bm R}_n}
\begin{bmatrix}
\hat \alpha_{{\bm k}s}\\
\hat \beta_{{\bm k}s}\\
\end{bmatrix}
\end{align}
with $N$ being the number of primitive cells in the graphene lattice and $\bm R_n$ being the $n$th Bravais lattice vector of the lattice, we can recast \Eq{graph} as
\be \label{h_graph_2}
H_{\gra}=-g\sum_{{\bm k}\in {\rm BZ}}\sum_s\Big(\theta_{\bm k}\hat\alpha_{{\bm k}s}^{\dg}\hat\beta_{{\bm k}s}+h.c.\Big),
\ee
where $\theta_{\bm k}\equiv \sum_{r=1}^{3} e^{i{\bm k}\cdot {\bm\Delta}_r}$ with the three $\Delta$-vectors (cf.\,\Fig{fig1}),
\[
\bm {\Delta}_1=\frac{a}{2}(\frac{1}{\sqrt{3}},1),\ \ \bm {\Delta}_2=\frac{a}{2}(\frac{1}{\sqrt{3}},-1),\ \ \bm {\Delta}_3=-a(\frac{1}{\sqrt{3}},0).
\]
Explicitly, $\theta_{\bm k}$ can be evaluated as
\begin{align}\label{theta}
\!\theta_{\bm k}
\!=\!\sqrt{\Big(\sin\frac{\sqrt{3}k_x a}{2}\Big)^2\!+\Big(\cos\frac{\sqrt{3}k_x a}{2}+2\cos\frac{k_y a}{2}\Big)^2},
\end{align}
a non-negative real number. Figure \ref{fig2} depicts the reciprocal lattice structure of graphene and the value of $\theta_{\bm k}$.
To diagonalize the Hamiltonian of \Eq{h_graph_2}, one may introduce
\bsube
\begin{align}
\hat c_{{\bm k}s}&=\frac{1}{\sqrt{2}}\Big(\hat\alpha_{{\bm k}s}-\hat\beta_{{\bm k}s}\Big),
\\
\hat d_{{\bm k}s}&=\frac{1}{\sqrt{2}}\Big(\hat\alpha_{{\bm k}s}+\hat \beta_{{\bm k}s}\Big),
\end{align}
\esube
and then obtain 
\be \label{hgra11}
H_{\gra}=\sum_{{\bm k} s}\varepsilon_{{\bm k}}\Big(\hat c_{{\bm k}s}^{\dg}\hat c_{{\bm k}s}-\hat d_{{\bm k}s}^{\dg}\hat d_{{\bm k}s}\Big)\ \ \text{with}\ \  \varepsilon_{{\bm k}}=g\theta_{\bm k}.
\ee

\subsection{Adatom-graphene hybridization}
Consider an atomic  impurity (adatom) absorbed on top of a carbon atom at site $n=0$ in sublattice $B$. 
The hybridization between the adatom and the graphene takes the form of \cite{Ynd14245420}
\be \label{hsb}
H_{\hyb}=-g_0\sum_{s}\big(\hat a_{s}^{\dg}\hat\beta_{0s}+\hat\beta_{0s}^{\dg} \hat a_{s}\big)+\epsilon_0\sum_{s}\hat\beta_{0 s}^{\dg}\hat\beta_{0s}.
\ee
Here, the former term represents the electron transfer coupling between the adatom and the graphene, while the latter term is the onsite energy modification due to the distortion. The distortion effects due to the impurity absorption are ignored in our theoretical model.\cite{Ynd14245420}  It is important to note that, although the onsite energy modification term is an operator within the pure graphene space, we treat it as an adatom-graphene hybridization with the system part being the system--space identity operator. This approach is justified for two reasons: Physically, the onsite energy modification results from the interaction between the adatom and the graphene; and, technically, the dispersion relation $\varepsilon_{\bm k} = g\theta_{\bm k}$ is challenging to derive analytically if we attempt to diagonalize both \Eq{graph} and the onsite energy modification term simultaneously.
That is if incorporating the term of onsite energy modification into the environment Hamiltonian, the onsite energy modification breaks the periodicity, making it hard to obtain the new environmental spectral density from the density of graphene.
Thus the extended dissipaton theory provides here a practical approach to handle substrates with local lattice symmertry breaking.

The influence of the graphene on the adatom is completely characterized by the hybridization spectral density  defined as
\begin{align} \label{spes}
\Gamma^{\sigma}_s(\omega) &\equiv \frac{1}{2}\int_{-\infty}^\infty\!\!\ud t\, e^{ -\sigma i\omega t}\la\{ \hat F^{\sigma}_s(t),\hat F^{\bar\sigma}_{s}(0) \}\ra_{\gra}.
\end{align}
Here, we denote 
\begin{align}
  \hat F_{s}^{-}\equiv \hat F_{s}\equiv -g_0\hat\beta_{0s}=\frac{g_0}{\sqrt{2N}}\sum_{{\bm k}\in {\rm BZ}}\big(\hat c_{{\bm k}s}-\hat d_{{\bm k}s}\big),
\end{align}
and $\hat F_{s}^{+}\equiv \hat F_s^{\dg}$, with $\sigma=\pm$ and $\bar\sigma$ being the sign opposite to that of $\sigma$ in \Eq{spes}.
We then have the fermionic fluctuation--dissipation theorem reading
\begin{align}\label{FDT}
 \la\hat F^{\sigma}_s(t)\hat F^{\bar\sigma}_{s'}(0)\ra_{\gra} &= \frac{\delta_{ss'}}{\pi}\int_{-\infty}^\infty\!\!\ud\omega\,\frac{ \Gamma_s (\omega)e^{\sigma i\omega t}}{1+e^{\sigma\beta\omega}}.
\end{align}
Here, the graphene spectral density is given by 
\begin{align}\label{spe}
\Gamma_s(\omega)=\frac{g_0^2\pi}{2N}\sum_{{\bm k}\in {\rm BZ}}\Big[\delta(\w-\varepsilon_{\bm k})+\delta(\w+\varepsilon_{\bm k})\Big].
\end{align}
It can be further evaluated as \cite{Cas09109,Ana1743705}
\begin{align}\label{spe222}
\Gamma_s(\omega)= \frac{g_0^2}{g} D\bigg( \frac{\omega^{}}{g} \bigg),
\end{align}
where the density of states
\begin{align}\label{D}
\!\!\!D(\zeta) = 
  \begin{cases}
      \dfrac{|\zeta|}{2\pi}\dfrac{1}{\sqrt{R(|\zeta|)}}\mathbb K\bigg( \dfrac{|\zeta|}{R(|\zeta|)} \bigg),\quad 0 \leq |\zeta| \leq 1,\\
      \dfrac{|\zeta|}{2\pi}\dfrac{1}{\sqrt{|\zeta|}}\mathbb K\bigg( \dfrac{R(|\zeta|)}{|\zeta|} \bigg),\quad 1 < |\zeta| \leq 3,
  \end{cases}
\end{align}
with 
\be\label{Rzeta}
R(\zeta) = (\zeta+1)^3(3-\zeta)/16
\ee
and the elliptic function of the first kind
\begin{align}\label{1stk}
  \mathbb K(x) \equiv  \int_0^1\frac{\ud z}{\sqrt{(1-z^2)(1-x z^2)}}.
\end{align}
We briefly summarized the derivations of \Eq{spe222}\cite{Cas09109,Ana1743705} in Appendix \ref{thappA}.

\subsection{Simulations on adatom-graphene composite}\label{num}

To apply the DEOM to adatom-graphene composite, we set
$H_{\tS}=H_{\ada}$ in \Eq{hs}.  
To proceed, we recast \Eq{hsb} as
\be 
H_{\hyb}=H_{\SB}^{\I}+H_{\SB}^{\rm\I\I}
\ee
where 
\begin{align} 
H_{\SB}^{\I}
=\sum_{s}\big(\hat a_{s}^{\dg}\ti F_{s}^{-}+\hat a_{s} \ti F_{s}^{+} \big),
\end{align}
and
\be 
H_{\SB}^{\I\I}=\epsilon_0\sum_{s}\hat\beta_{0 s}^{\dg}\hat\beta_{0s}=
\frac{\epsilon_0}{2g_0^2}\sum_{s,\sigma\sigma'}\hat q^{\bar\sigma \bar \sigma'}_{s}\ti F_s^{\sigma}\ti F_s^{\sigma'}.
\ee
Here, $\ti F^{-}_{s}\equiv-g_0\hat\beta_{0s}$ and $\ti F^{+}_{s}\equiv g_0\hat\beta_{0s}^{\dg}$, whereas $\hat q^{-+}_{s}=-\hat I$, $\hat q^{+-}_{s}=\hat I$ and $\hat q^{++}_{s}=\hat q^{--}_{s}=0$ for any $s$, with $\hat I$ being the identity operator in the reduced system space.
The simulation results are exhibited with the adatom spectral density
\begin{equation}\label{As}
	A_{s}(\w) = \frac{1}{2 \pi} \!\int_{-\infty}^{\infty}\!\!{\rm d} t\, e^{i\w t} \la\{ \hat a_{s} (t), \hat a_{s}^{\dagger} (0)\} \ra.
\end{equation}
The details of evaluating the correlation functions have been elabrated in \Sec{subsection:2.4}.

\begin{figure}[ht]  
\centering\includegraphics[width=0.85\columnwidth]{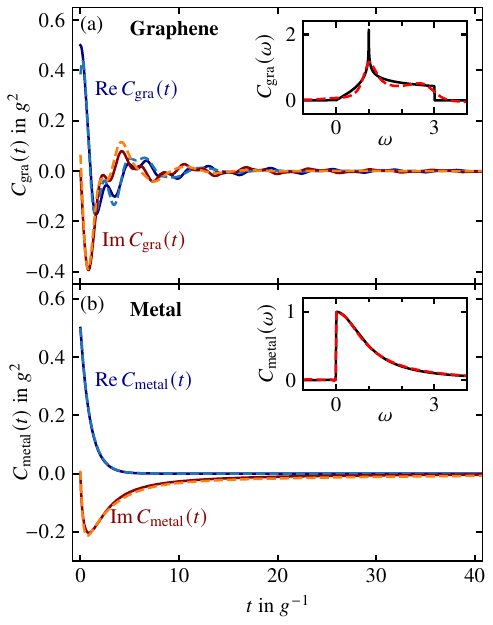} 
\caption{The hybridization bath   correlation functions, \Eq{FDT} at $T=77$K, with $\Gamma_s(\omega)$ as \Eq{spe222} of graphene (upper-panel) and the corresponding \Eq{Lorentz} of metal (lower-panel): Exact
(solid) versus the $t$-PFD results (dashed) and the frequency-domain counterparts shown in the insets where the units of both axes are $g$; see text for more details.}    
\label{fig3}   
\end{figure}

For numerical demonstrations, we set $g=2.8\,$eV in \Eq{graph} for the nearest-neighboring electron transfer coupling in graphene.\cite{Cas09109} Besides, we select $T= 0.0024\,g=77\,\mathrm K$ and $g_0=g$ in \Eq{hsb} for the adatom-graphene coupling. 
The hybridization spectral density in \Eq{spe222} is then determined. Depicted in the upper panel of \Fig{fig3} are $C_{\rm gra}(t)\equiv \la\hat F^{-}_{s}(t)\hat F^{+}_{s}(0)\ra_{\rm gra}$  and its spectrum $C_{\rm gra}(\w)$ of graphene, together with the results of time-domain Prony fitting decomposition ($t$-PFD) scheme.\cite{Che22221102} Given the computational constraints, we use 8 exponential terms to decompose $C_{\rm gra}(t)$ for the extended DEOM simulations. For later comparisons between graphene and metal substrates, we depict the corresponding results of metal in the lower panel of \Fig{fig3} as well. For the metal, the hybridization environmental spectral density adopts the Lorentzian form,
\begin{align}\label{Lorentz}
\Gamma_s^{\rm metal}(\omega)= \frac{g_0^2}{g} \frac{ W^2}{\omega^2 + W^2}\ \ \ \ (\text{Lorentzian}),
\end{align}
for both $s=\,\uparrow$ and $\downarrow$. We select the bandwidth $W=g$ and use 5 exponential terms to fit the time domain correlation function $C_{\rm metal}(t)$.

\begin{figure}[t]  
\centering\includegraphics[width=0.85\columnwidth]{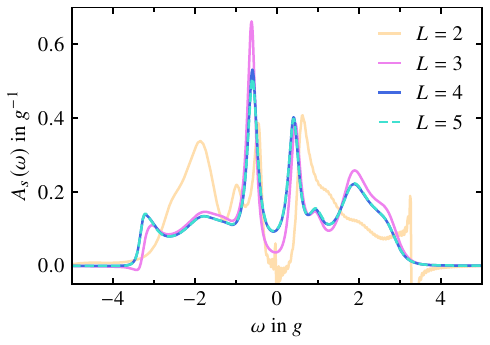} 
\caption{The simulation results of $A_s(\w)$ with $U=2.8g=2|\epsilon_s|$, $\epsilon_0=-0.7g$, and varied
truncation levels as labeled.}
\label{fig4}   
\end{figure}

Figure \ref{fig4} presents the simulation results of $A_s(\w)$ with $U = 2|\epsilon_s|$. Other parameters are $\epsilon_{\uparrow}=\epsilon_{\downarrow}=-1.4g$ in \Eq{hs} for the adatom and $\epsilon_0=-0.7g$ in \Eq{hsb} for the onsite energy modification of carbon.\cite{Sof12115405} The truncation tier levels of $L$ are set from 2 to 5. The case of $L=2$ roughly corresponds to the self-consistent quantum master equation level including the quadratic system-bath coupling, which is obviously insufficient. The truncation with $L=4$ provides numerically reliable results.
%\cite{Li12266403, Han18234108,Din22224107}

\begin{figure*}[ht]  
\centering\includegraphics[width=1\textwidth]{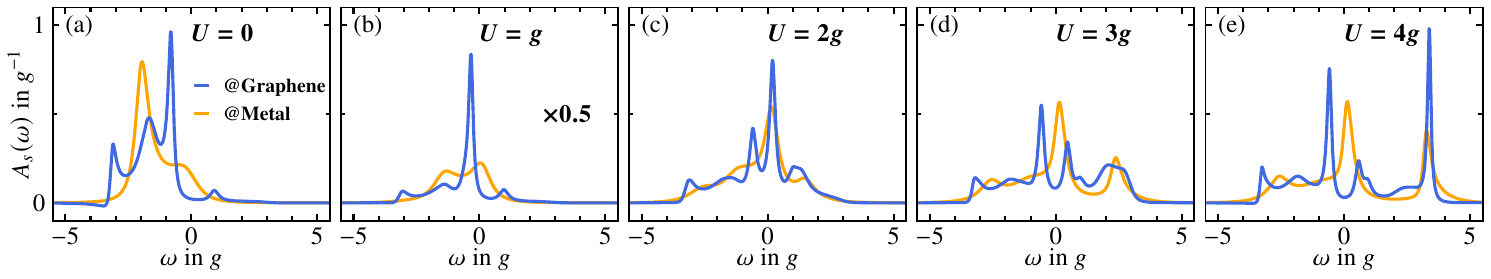} 
\caption{The simulation results of $A_s(\w)$ with varied values of $U$ for the adatoms at graphene (blue curves). 
For comparison, we also simulate the results of the adatom at a model metal substrate 
(yellow curves).}    
\label{fig5}   
\end{figure*}

\begin{figure}[ht]  
\centering\includegraphics[width=0.85\columnwidth]{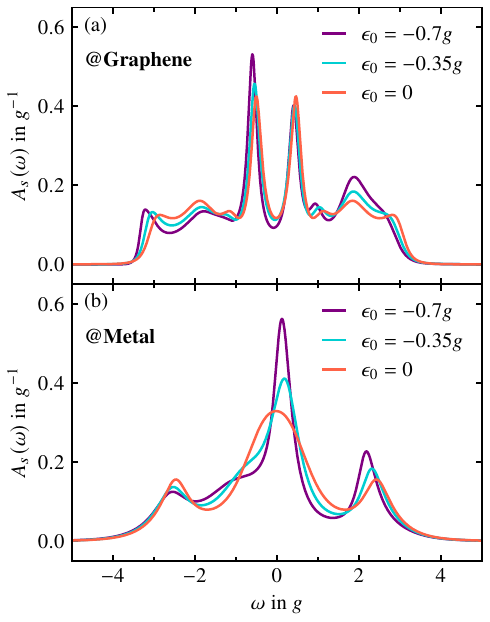} 
\caption{The simulation results of $A_s(\w)$ with $U=2.8g$ and varied values of onsite energy modification strength $\epsilon_0$. See the text for details. }
\label{fig6}   
\end{figure}

\begin{figure}[ht] 
\centering\includegraphics[width=0.85\columnwidth]{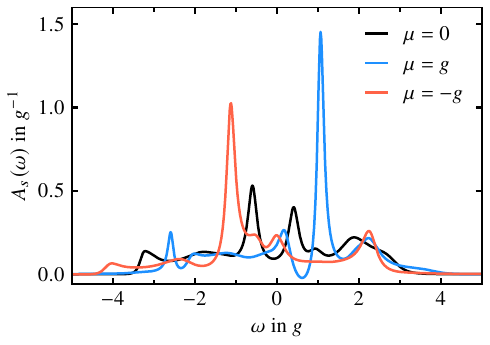} 
\caption{The simulation results of $A_s(\w)$ with $U=2.8g$, $\epsilon_0 = -0.7g$, and varied values of voltage $\mu$ applied on the graphene. See the text for details.}
\label{fig7}   
\end{figure}

Figure \ref{fig5} depicts the simulation results of $A_s(\w)$ with $\epsilon_{\uparrow}=\epsilon_{\downarrow}=-1.4g$, $\epsilon_0=-0.7g$, and varied values of $U$. The blue curves represent the results of the adatoms at graphene, while the yellow curves correspond to those of the adatoms at the metal substrate for comparison. In the panel (a) of \Fig{fig5}, the emergence of non-positive spectrum around $\omega = -3.3g$ (negative part is also seen in the blue curve of \Fig{fig7}) is due to the numerical error in exponential decomposition of the environment correlation $C_{\rm gra}(t)$ (see \Fig{fig3}). For the metal environment with Lorentzian spectral densities, the Hubbard peaks are around $\epsilon_{s}-\lambda$ and $\epsilon_{s}-\lambda+U$, where $\lambda=g_0^2/g$ denotes the reorganization energy. For the present parameters, the Hubbard peaks will appear around $-2.4g$ and $-2.4g+U$. 
These peaks indeed exist in the yellow curves, but exhibit manifest shifts due to the onsite energy modification; cf.\,the second term in \Eq{hsb} and the analysis after \Fig{fig6}. 
For the metal substrate, when $U$ is about or larger than $-2\epsilon_s=2.8g$, the Kondo peak emerges at $\w=0$, originating from strong electronic correlation effects.\cite{Phi12}
Compared to those of the yellow curves, the peaks of blue curves  
exhibit intricate behaviors for the adatom at the graphene substrate reflecting the band structure of graphene (cf.\ the two peaks in the inset of the upper panel of \Fig{fig3}), with the Kondo mechanism also playing roles. For further analysis, see \Figs{fig6} and \ref{fig7} including the remarks therein.

In \Fig{fig6}, we demonstrate the effects of the onsite energy modification term, namely, the quadratic coupling in the adatom--substrate hybridization model. The results are simulated under the particle-hole symmetric point\cite{Phi12} $U = 2|\epsilon_s|$, with varied values of the modification strength $\epsilon_0$. By comparing the results of $\epsilon_0 \neq 0$ to those of $\epsilon_0 = 0$ for both the graphene environment (upper panel) and the metal environment (lower panel), we observe that the Kondo effect primarily arises from the linear interaction ($\epsilon_0 = 0$). When the bath interacts with the system purely linearly ($\epsilon_0 = 0$), the adatom spectral density on a metal substrate displays a prominent Kondo peak near the Fermi level ($\omega = 0$) and two Hubbard peaks around $\omega = \pm U$. For the graphene case with $\epsilon_0 = 0$, the spectral density has a similar behavior with the metal one, but the three main peaks are split and reshaped due to graphene’s intricate band structure (cf.\ the two peaks in the inset of the upper panel of \Fig{fig3}). Introducing the quadratic couplings (onsite energy modifications) not only shifts the positions of each peak but also alters their intensities. More importantly, this quadratic interaction breaks the particle-hole symmetry within the system–bath hybridization dynamics, leading to asymmetric spectral densities, i.e., $A_s(\omega) \neq A_{s}(-\omega)$ when $\epsilon_0$ is nonzero. Consequently, for an adatom--graphene system, the spectral function exhibits more complex peak structures induced by strongly correlated many-body interactions, as shown in \Fig{fig5}.

To the end of this section, we discuss the Kondo mechanism caused by the graphene substrate. 
The Kondo scattering refer to the spin-exchange induced screening effect caused by the strong electronic correlation, most efficiently occurred near the Fermi level.\cite{Phi12} However, in graphene, the density of states near the Fermi level $\omega = 0$ is very low. To illustrate the Kondo peaks, we can apply an external voltage to shift the electronic states so that electrons at the peak of the spectral density (cf.\ the inset of the upper panel of \Fig{fig3}) could participate in Kondo scattering. The  results are shown in \Fig{fig7}.
Originally ($\mu=0$), the adatom spectral density exhibits no Kondo peak at the Fermi level, as the Fermi level  is located at $\omega = 0$, where the spectral density vanishes (inset of the upper panel of \Fig{fig3}). When an external voltage is applied on the graphene, the Fermi level effectively shifts. To illustrate this effect, \Fig{fig7} presents the adatom spectral densities under the particle-hole symmetric point $U = 2|\epsilon_s|$ for varied values of applied voltage $\mu$. In the presence of a finite voltage, the exponential decomposition of the bath correlation function, cf.\,\Eq{Fc}, is modified as
$\gamma_{\kappa s}^\sigma \to \gamma_{\kappa s}^\sigma - \sigma i\mu $ 
while $\{\eta_{\kappa s}^\sigma\}$ remain unchanged.\cite{Ye16608} As shown in \Fig{fig7}, the Kondo peak emerges sharply around $\omega = \mu$ with $\mu = g$ (blue line) and $\mu = -g$ (red line), cf.\ the inset of the upper panel of \Fig{fig3}.

\section{Summary}\label{thsum}

To summarize, in this paper we systematically develop in general the extended dissipaton-equation-of-motion (DEOM) and its equivalent dissipaton-embedded quantum master equation (DQME). 
Both the extended DEOM and DQME are exact for open quantum systems interacting with environments composed of non-interacting electrons, allowing for the accurate treatment of both linear and quadratic environmental couplings.

The DEOM is identical to the well-established HEOM formalism when only the linear system-bath coupling is involved and reduced system dynamics is considered. 
The HEOM is rooted from the  Feynman--Vernon influence functional path integral.
All numerical methods developed for HEOM are applicable in DEOM evaluations. 
However, in the dissipaton theory, the underlying dissipaton algebra can be  straightforwardly applied to collective bath dynamics and nonlinear bath couplings,
enable the extension of theory readily.
On the other hand, the  DQME supplies the dissipaton a  more concrete quasi-particle picture. 
Its formulation as a single master equation, not hierarchical equations,  enhances its suitability and convenience for quantum algorithms \cite{Li24032620} and other advanced computational methods .

As a practical application, the extended dissipaton theory is used to study the strongly correlated properties of an adatom on graphene by simulating the adatom's spectral functions and comparing to those at the metal substrate. 
The present theory handles both linear and quadratic environmental couplings which account for the interactions between the adatom and the substrate.
We carry out numerical simulations on the adatom spectral function with varied parameters under the influence of different substrate band characters and analyze the underlying mechanism,
with highlighting
the strong electronic correlation effects and Kondo features.
It is worth reemphasizing that the extended dissipaton theory presented in this paper provides universal methods
in treating strongly correlated open  systems.

\vspace{1em}
Support from the Ministry of Science and Technology of China (Grant No.\ 2021YFA1200103), the National Natural Science Foundation of China (Grant Nos.\  22173088, 22321003, 22373091, 22393912, 22425301, 224B2305), the Strategic Priority Research Program of Chinese Academy of Sciences (Grant No.\ XDB0450101), and the Innovation Program for Quantum Science and Technology (Grant No.\ 2021ZD0303301) is gratefully acknowledged.

\appendix

\section{Derivation of extended DEOM \Eq{DEOM_wy0828}}\label{DEOM_derivation}
\subsection{Dissipaton algebra with extended Wick's theorem}

To complete the dissipaton theory, 
we  introduce the dissipaton algebra composed of two important ingredients: ($i$) Each dissipaton satisfies the generalized diffusion equation,
\begin{align}\label{gdiff}
  {\rm tr}_{\B}\bigg[ \bigg( \frac{\partial}{\partial t}\hat f_j \bigg)_{\B}\rho_{\T}(t) \bigg] = -\gamma_j{\rm tr}_{\B}[\hat f_j\rho_{\T}(t)],
\end{align}
where $(\frac{\partial}{\partial t}\hat f_j)_{\B} = -i[\hat f_j,h_{\B}]$. 
Equation (\ref{gdiff})
arises from that each dissipaton is associated with a single exponent, for its forward and backward correlation functions [cf.\,\Eq{fcf}]; ($ii$) The generalized Wick's theorems (GWTs) deal with adding dissipaton operators into the irreducible notation. The GWT-1 evaluates the linear bath coupling with one dissipaton added each time, expressed as 
\begin{subequations}\label{GWT1}
  \begin{align}
    &\quad\,{\rm tr}_{\B}\Big[ (\hat f_{j_n}\cdots\hat f_{j_1})^\circ\hat f_j\rho_{\T}(t) \Big]\nl
    & = (-)^n\bigg[\rho_{{\bf j}j}^{(n+1)} - \sum_{r=1}^n(-)^{n-r}\la\hat f_{j_r}\hat f_j\ra^>_{\B}\rho_{{\bf j}_r^-}^{(n-1)}\bigg],\\
    &\quad\,{\rm tr}_{\B}\Big[ \hat f_j(\hat f_{j_n}\cdots\hat f_{j_1})^\circ\rho_{\T}(t) \Big]\nl
    & = \rho_{{\bf j}j}^{(n+1)}+\sum_{r=1}^n(-)^{n-r}\la\hat f_j\hat f_{j_r}\ra^<_{\B}\rho_{{\bf j}_r^-}^{(n-1)},
  \end{align}
\end{subequations}
where $\la\hat f_{j}\hat f_{j'}\ra^{\lessgtr}_{\B}$ is as defined below \Eq{fcf}, ${\bf j} j\equiv \{j_1\cdots j_nj\}$, and ${\bf j}_r^- \equiv \{ j_1 \cdots j_{r-1}j_{r+1}\cdots j_n\}$. Moreover, the GWT-2 is concerned with the environmental quadratic  couplings, where two dissipatons act simultaneously.
It reads
\bsube\label{GWT2}
  \begin{align}
    &\quad\,{\rm tr}_{\B}[(\hat f_{j_n}\cdots\hat f_{j_1})^\circ\hat f_j\hat f_{j'}\rho_{\T}]\nl 
  &= \rho_{{\bf j}j'j}^{(n+2)}+\la\hat f_j\hat f_{j'}\ra_{\B}\rho_{\bf j}^{(n)}-\sum_{r=1}^n(-)^{n-r}\la\hat f_{j_r}\hat f_{j'}\ra_{\B}^>\rho_{{{\bf j}}_{r}^-j}^{(n)}\nl 
  &\quad +\sum_{r=1}^n(-)^{n-r}\la\hat f_{j_r}\hat f_j\ra^>_{\B}\rho_{{\bf j}_r^-j'}^{(n)}\nl 
  &\quad +\sum_{r,r'}(-)^{r-r'+\Theta(r'-r)}\la\hat f_{j_r}\hat f_j\ra_{\B}^>\la\hat f_{j_{r'}}\hat f_{j'}\ra_{\B}^>\rho_{{\bf j}_{rr'}^{--}}^{(n-2)},
\end{align}
and
\begin{align}
&\quad {\rm tr}_{\B}[\hat f_j\hat f_{j'}(\hat f_{j_n}\cdots\hat f_{j_1})^\circ \rho_{\T}]\nl 
  &= \rho_{{\bf j}j'j}^{(n+2)}+\la\hat f_j\hat f_{j'}\ra_{\B}\rho_{\bf j}^{(n)}-\sum_{r=1}^n(-)^{n-r}\la\hat f_{j}\hat f_{j_r}\ra_{\B}^<\rho_{{\bf j}_{r}^-{j'}}^{(n)}\nl 
  &\quad +\sum_{r=1}^n(-)^{n-r}\la\hat f_{j'}\hat f_{j_r}\ra^<_{\B}\rho_{{\bf j}_r^-j}^{(n)}\nl 
  &\quad -\sum_{rr'}(-)^{r-r'+\Theta(r'-r)}\la\hat f_{j'}\hat f_{j_r}\ra_{\B}^<\la\hat f_j\hat f_{j_{r'}}\ra_{\B}^<\rho_{{\bf j}_{rr'}^{--}}^{(n-2)}.
  \end{align}
\esube
Here, $\Theta(x)$ is the Heviside step function 
\[
\Theta(x) =
\begin{cases}
1 &\quad\text{if}\ \  x\geq 0, \\
0 & \quad\text{if}\ \  x<0,
\end{cases}
\]
and 
\[
{\bf j}_{rr'}^{--} \equiv \{ j_1\cdots j_{r-1}j_{r+1}\cdots j_{r'-1}j_{r'+1} \cdots j_n\} = {\bf j}_{r'r}^{--}.
\]

\subsection{Extended DEOM derivation}

By applying the dissipaton algebra on the von Neumann--Liouville equation,
\begin{align}
  \dot{\rho}_{\T} = -i[H_{\T},\rho_{\T}] = -i[H_{\tS}+h_{\B}+H_{\SB}^{\I}+H_{\SB}^{\I\I},\rho_{\T}],
\end{align}
one can construct the extended DEOM. We now elaborate, term by term, the contributions of specific four components in $H_{\T}$.
\begin{enumerate}[fullwidth,label=(\emph{\rm\alph*})]
  \item The $H_{\tS}$--contribution: Evidently, 
  \begin{align}
    {\rm tr}_{\B}\{(\hat f_{j_n}\cdots\hat f_{j_1})^\circ 
    [H_{\tS},\rho_{\T}]\} = [H_{\tS},\rho_{\bf j}^{(n)}].
  \end{align}
  
  \item The $h_{\B}$--contribution: Using \Eq{gdiff}, we have 
  \begin{align}
    i{\rm tr}_{\B}\Big\{ (\hat f_{j_n}\cdots\hat f_{j_1})^\circ [h_{\B},\rho_{\T}] \Big\} = \sum_{r=1}^n\gamma_{j_r}\rho_{\bf j}^{(n)}.
  \end{align}
  % with $\gamma^{(n)}_{\bf j}\equiv \sum_{r=1}^n\gamma_{j_r}$.
  \item The $H_{\SB}^{\I}$--contribution: By using \Eqs{SB1f} and (\ref{GWT1}), we evaluate 
  \bsube
  \begin{align}
    &\sum_j{\rm tr}_{\B}[(\hat f_{j_n}\cdots\hat f_{j_1})^\circ\hat a_{\bar j}\hat f_j\rho_{\T}]\nl
    =& \sum_j(-)^n\hat a_{\bar j}{\rm tr}_{\B}[(\hat f_{j_n}\cdots\hat f_{j_1})^\circ\hat f_j\rho_{\T}]\nl 
    =& \sum_j\hat a_{\bar j}\rho_{{\bf j}j}^{(n+1)}+\sum_{r=1}^n\sum_v(-)^{n-r}\eta^{\sigma_r}_{\kappa_ru_rv}\hat a^{\sigma_r}_{v}\rho_{{\bf j}_r^-}^{(n-1)},
  \end{align}
  and 
  \begin{align}
    &\sum_j{\rm tr}_{\B}[(\hat f_{j_n}\cdots\hat f_{j_1})^\circ\rho_{\T}\hat a_{\bar j}\hat f_j]\nl 
    =&\sum_j (-)^n{\rm tr}_{\B}[\hat f_j(\hat f_{j_n}\cdots\hat f_{j_1})^\circ\rho_{\T}]\hat a_{\bar j}\nl 
    =&(-)^{n}\bigg[\sum_j\rho_{{\bf j}j}^{(n+1)}\hat a_{\bar j}-\sum_{r=1}^n\sum_v(-)^{n-r}\eta^{\bar\sigma_r*}_{\kappa_ru_rv}\rho_{{\bf j}_r^-}^{(n-1)}\hat a^{\sigma_r}_{v}\bigg].
  \end{align}
  \esube
  \item The $H_{\SB}^{\I\I}$--contribution: By applying \Eqs{SB2f} and (\ref{GWT2}), we can readily obtain
  \begin{align}\label{wy828}
    &\quad {\rm tr}_{\B}\Big\{ (\hat f_{j_n}\cdots\hat f_{j_1})^\circ [H_{\SB}^{\I\I},\rho_{\T}] \Big\}\nl 
    &= \frac{1}{2}\sum_{jj'}\hat q_{\bar j\bar j'}{\rm tr}_{\B}\Big[ (\hat f_{j_n}\cdots\hat f_{j_1})^\circ \hat f_j\hat f_{j'}\rho_{\T} \Big]\nl &
    \quad -\frac{1}{2}\sum_{jj'}{\rm tr}_{\B}\Big[ (\hat f_{j_n}\cdots\hat f_{j_1})^\circ \rho_{\T}\hat f_j\hat f_{j'} \Big]\hat q_{\bar j\bar j'}\nl
    &= \frac{1}{2}\sum_{jj'}[\hat q_{\bar j\bar j'},\rho_{{\bf j}j'j}^{(n+2)}]+\frac{1}{2}\sum_{\sigma us}\sum_{\sigma' vs'}\la\tilde F^\sigma_{us}\tilde F^{\sigma'}_{vs'}\ra_{\B}[\hat q^{\bar\sigma\bar\sigma'}_{us,vs'},\rho_{\bf j}^{(n)}]\nl 
    &\quad -\sum_{r=1}^n\sum_{vj}(-)^{n-r}\Big[ \eta^{\sigma_r}_{\kappa_ru_rvs_r}\hat{q}^{\sigma_r\bar\sigma}_{vs_r,us}\rho_{{\bf j}_r^-j}^{(n)}\nl 
    &\qquad\qquad\qquad\quad\qquad+\eta^{\bar\sigma_r*}_{\kappa_ru_rvs_r}\rho_{{\bf j}_r^-j}^{(n)}\hat{q}^{\sigma_r\bar\sigma}_{vs_r,us} \Big]\nl 
    &\quad +\sum_{r>r'}\sum_{uv}(-)^{r-r'}\Big[ \eta^{\sigma_r}_{\kappa_ru_rus_r}\eta^{\sigma_{r'}}_{\kappa_{r'}u_{r'}vs_{r'}}\hat{q}^{\sigma_r\sigma_{r'}}_{us_r,vs_{r'}}\rho_{{\bf j}_{rr'}^{--}}^{(n-2)}\nl 
    &\qquad -\eta^{\bar\sigma_r*}_{\kappa_ru_rus_r}\eta^{\bar\sigma_{r'}*}_{\kappa_{r'}u_{r'}vs_{r'}}\rho_{{\bf j}_{rr'}^{--}}^{(n-2)}\hat{q}^{\sigma_r\sigma_{r'}}_{us_r,vs_{r'}}\Big].
  \end{align}
\end{enumerate}
Therefore, we obtain the final and full formalism of the extended DEOM as in \Eq{DEOM_wy0828} with \Eqs{eq19}--(\ref{eq23}).

\section{Derivation of extended DQME \Eq{eDQME}}\label{appB}

To derive \Eq{eDQME}, we firstly rewrite the extended DEOM [\Eq{DEOM_wy0828}] in the dissipaton occupation number representation. That is, we relabel the dissipaton density operators as 
\begin{align}
    \rho_{\bf j}^{(n)} \longmapsto \rho_{\mathbf{nm}} \equiv \rho_{n_1n_2\cdots n_{\mathcal K}; m_1m_2\cdots m_{\mathcal K}},
\end{align}
where $n_k,m_k = 0,1$ are the occupation numbers of the dissipatons $\hat f_j$ with $j = (\sigma\kappa us)\big|_{\sigma = +} \equiv (+,k)$
and $j =(\sigma\kappa us)\big|_{\sigma = -} \equiv (-,k)$, respectively. $\mathcal K$ is the half of total number of involved dissipatons. In terms of the new labels, we have 
\begin{align}
    \begin{split}
        \rho_{\mathbf j;(+,k)}^{(n+1)} &\longmapsto (-)^{M+N-\theta_k^+}\rho_{\mathbf n_k^+\mathbf m}, \\
        \rho_{\mathbf j;(-,k)}^{(n+1)} &\longmapsto (-)^{M-\theta_k^-}\rho_{\mathbf n\mathbf m_k^+},
    \end{split}
\end{align}
where $\theta_k^+\equiv\sum_{l=1}^kn_l$, $\theta_k^-\equiv\sum_{l=1}^km_l$,  $M \equiv \theta_{\mathcal K}^+$, and $N\equiv \theta_{\mathcal K}^-$.
Denote $\gamma_{\bf nm} \equiv \sum_{k}(n_k\gamma^+_k + m_k\gamma^-_k)$.
Consequently, the extended DEOM [\Eq{DEOM_wy0828}] is recasted as
\begin{widetext}
  \begin{align}\label{rhonm}
    \dot\rho_{\bf nm} &= -(i\mathcal L_{\tS}^{\rm eff} + \gamma_{\bf nm})\rho_{\bf nm} \nl
    &\quad - i\sum_{k}\Big[ (-)^{M+N-\theta_k^+}\hat a_k^-\rho_{\mathbf n_k^+\mathbf m} - (-)^{\theta_k^+}\rho_{\mathbf n_k^+\mathbf m}\hat a_k^- \nl
    &\quad\quad\qquad + (-)^{M-\theta_k^-}\hat a_k^+\rho_{\mathbf{nm}_k^+} -(-)^{N+\theta_k^-}\rho_{\mathbf{nm}_k^+}\hat a_k^+\Big] \nl
    &\quad - i \sum_{k}\Big[ (-)^{M+N-\theta_k^+}\eta_{k}^+\hat a_{k}^+\rho_{\mathbf n_k^-\mathbf m} + (-)^{\theta_k^+}\eta_{k}^{-*}\rho_{\mathbf n_k^-\mathbf m}\hat a_{k}^+ \nl
    &\quad \quad\qquad+(-)^{M-\theta_k^-}\eta_{k}^-\hat a_{k}^-\rho_{\mathbf{nm}_k^-} + (-)^{N+\theta_k^-}\eta_{k}^{+*}\rho_{\mathbf{nm}_k^-}\hat a_{k}^-\Big]\nl
    &\quad -i\!\sum_{k>k'}\!\bigg\{(-)^{\theta_k^+-\theta_{k'}^+}\Big[ \hat q^{--}_{kk'},\rho_{\mathbf n_{kk'}^{++}\mathbf m} \Big]
   % \nl &\quad
    \!+\! (-)^{\theta_k^--\theta_{k'}^-}\Big[ \hat q^{++}_{kk'},\rho_{\mathbf n\mathbf m_{kk'}^{++}} \Big]  \bigg\}\nl
    &\quad-i\sum_{kk'}(-)^{N+\theta_k^--\theta_{k'}^+}\Big[ \hat q^{+-}_{kk'},\rho_{\mathbf n_{k'}^{+}\mathbf m_{k}^{+}} \Big]\nl
    &\quad-i \sum_{k>k'}\bigg[(-)^{\theta_k^+-\theta_{k'}^+}\Big( \eta^+_k\eta^+_{k'}\hat q^{++}_{kk'}\rho_{\mathbf n_{kk'}^{--}\mathbf m} - \eta_{k}^{-*}\eta_{k'}^{-*}\rho_{\mathbf n_{kk'}^{--}\mathbf m} \hat q^{++}_{kk'}\Big)\nl
    &\quad\ \ \ \qquad + (-)^{\theta_k^--\theta_{k'}^-}\Big( \eta^-_k\eta^-_{k'}\hat q^{--}_{kk'}\rho_{\mathbf n\mathbf m_{kk'}^{--}} - \eta_{k}^{+*}\eta_{k'}^{+*}\rho_{\mathbf n\mathbf m_{kk'}^{--}} \hat q^{--}_{kk'}\Big)  \bigg]\nl
    &\quad-i\sum_{kk'}(-)^{N+\theta_k^--\theta_{k'}^+}\Big( \eta^-_{k}\eta^{+}_{k'}\hat q^{-+}_{kk'}\rho_{\mathbf n_{k'}^-\mathbf m_k^-} - \eta^{+*}_{k}\eta^{-*}_{k'}\rho_{\mathbf n_{k'}^-\mathbf m_k^-}\hat q^{-+}_{kk'}\Big)\nl
    &\quad-i\!\sum_{k k'}(-)^{\Theta(k-k')}\!\bigg[\!(-)^{\theta_k^+-\theta_{k'}^+}\Big( \eta^+_{k'}\hat q^{+-}_{k'k}\rho_{\mathbf n_{kk'}^{+-}\mathbf m} \!+ \eta^{-*}_{k'}\rho_{\mathbf n_{kk'}^{+-}\mathbf m}\hat q^{+-}_{k'k} \Big)\nl
    &\quad\quad\qquad + (-)^{\theta_k^--\theta_{k'}^-}\Big( \eta^-_{k'}\hat q^{-+}_{k'k}\rho_{\mathbf n\mathbf m_{kk'}^{+-}} + \eta^{+*}_{k'}\rho_{\mathbf n\mathbf m_{kk'}^{+-}}\hat q^{-+}_{k'k} \Big) \bigg]\nl
    &\quad-i\sum_{kk'}(-)^{N+\theta_k^+-\theta_{k'}^-}\Big( \eta^-_{k'}\hat q^{--}_{k'k}\rho_{\mathbf n^-_{k}\mathbf m^+_{k'}} + \eta^{+*}_{k'}\rho_{\mathbf n^-_{k}\mathbf m^+_{k'}}\hat q^{--}_{k'k} \Big)
    \nl
   &\quad+i\sum_{kk'}(-)^{N+\theta_k^--\theta_{k'}^+}\Big( \eta^+_{k'}\hat q^{++}_{k'k}\rho_{\mathbf n^-_{k'}\mathbf m^+_{k}} + \eta^{-*}_{k'}\rho_{\mathbf n^-_{k'}\mathbf m^+_{k}}\hat q^{++}_{k'k} \Big) 
    .
\end{align}

To proceed, we introduce the fermionic creation and annihilation operators $\{\hat b_k^\sigma\}$ and the corresponding particle-number basis, 
\begin{align}
    \begin{split}
        |\mathbf m\ra &\equiv (\hat b_1^+)^{m_1}\cdots (\hat b_{\mathcal K}^+)^{m_{\mathcal K}} |\mathbf 0\ra, \\
        |\mathbf n\ra &\equiv (\hat b_{\mathcal K}^+)^{n_{\mathcal K}}\cdots  (\hat b_1^+)^{n_1}|\mathbf 0\ra.
    \end{split}
\end{align}
Substituting \Eq{rhonm} into 
\begin{align}
    \tilde\rho =  \sum_{\mathbf{mn}}|\mathbf m\ra\bar\rho_{\mathbf{nm}}\la\mathbf n|
\end{align}
with 
\begin{align}
    \bar\rho_{\mathbf{nm}} = \prod_{kk'}\frac{1}{(\zeta_k^-)^{m_k}(\zeta_{k'}^+)^{n_{k'}}}\rho_{\mathbf {nm}},
\end{align}
and noticing the following identities,
\be 
\begin{split}
        &\hat b_k^+|\mathbf m\ra = (-)^{\theta_k^-}|\mathbf m_k^+\ra,\quad\ \ \ \,
        \hat b_k^-|\mathbf m\ra = (-)^{\theta_k^--1}|\mathbf m_k^-\ra,\\
        &\hat b_k^+|\mathbf n\ra = (-)^{N-\theta_k^+}|\mathbf m_k^+\ra,\quad
        \hat b_k^-|\mathbf n\ra = (-)^{N-\theta_k^+}|\mathbf n_k^-\ra,
        \end{split}
\ee
we obtain the extended DQME, 
\begin{align}\label{eDQME0}
    \dot{\tilde\rho} &= -i\mathcal L_{\tS}^{\rm eff}\tilde\rho - \sum_k(\gamma_k^-\hat N_k\tilde\rho + \gamma_k^+\tilde\rho\hat N_k) \nl
    &\quad\,-i \sum_k\Big[ \zeta_k^-\big( \hat a_k^+\hat b_k^-\tilde\rho - \hat b_k^-\tilde\rho\hat a_k^+ \big) + \zeta_k^+\big( \hat a_k^-\tilde\rho\hat b_k^+ - \tilde\rho\hat b_k^+\hat a_k^- \big)\nl
    &\quad\quad + \xi_k^+\hat a_k^+\tilde\rho\hat b_k^- + \xi_k^{-*}\tilde\rho\hat b_k^-\hat a_k^+ - \xi_k^-\hat a_k^-\hat b_k^+\tilde\rho - \xi_k^{+*}\hat b_k^+\tilde\rho\hat a_k^- \Big] \nl
    &\quad\, -\frac{i}{2}\sum_{kk'}\Big\{ \zeta_k^-\zeta_{k'}^-[\hat q_{kk'}^{--},\tilde\rho\hat b_k^+\hat b_{k'}^+] + \zeta_k^+\zeta_{k'}^+[\hat q_{kk'}^{++}, \hat b_k^-\hat b_{k'}^-\tilde\rho]\nl
    &\quad\quad + \zeta_{k}^-\zeta_{k'}^+[\hat q^{+-}_{kk'},\hat b_k^-\tilde\rho\hat b_{k'}^+] - \zeta_k^+\zeta_{k'}^-[\hat q^{-+}_{kk'}, \hat b^-_{k'}\tilde\rho\hat b_k^+]\nl
    &\quad\quad + \xi_k^+\xi_{k'}^+\hat q_{kk'}^{++}\tilde\rho\hat b_{k}^-\hat b_{k'}^- - \xi_k^{-*}\xi_{k'}^{-*}\tilde\rho\hat b_{k}^-\hat b_{k'}^-\hat q_{kk'}^{++} \nl
    &\quad\quad + \xi_k^-\xi_{k'}^-\hat q^{--}_{kk'}\hat b_k^+\hat b_{k'}^+\tilde\rho - \xi_k^{+*}\xi_{k'}^{+*}\hat b_k^+\hat b_{k'}^+\tilde\rho\hat q^{--}_{kk'} \nl
    &\quad\quad -\xi_k^-\xi_{k'}^+\hat q^{-+}_{kk'}\hat b_k^+\tilde\rho\hat b_{k'}^- + \xi_k^{+*}\xi_{k'}^{-*}\hat b_k^+\tilde\rho\hat b_{k'}^-\hat q^{-+}_{kk'}\nl
    &\quad\quad + \xi_k^+\xi_{k'}^-\hat q_{kk'}^{+-}\hat b_{k'}^+\tilde\rho\hat b_k^- - \xi_k^{-*}\xi_{k'}^{+*}\hat b_{k'}^+\tilde\rho\hat b_k^-\hat q_{kk'}^{+-} \Big\} \nl
    &\quad\, -i \sum_{kk'}\Big[ \zeta_k^-\big( \xi_{k'}^+\hat q_{k'k}^{+-}\tilde\rho\hat b^-_{k'}\hat b^+_{k} + \xi_{k'}^{-*}\tilde\rho\hat b^-_{k'}\hat b^+_{k}\hat q_{k'k}^{+-} \big) \nl
    &\quad\quad + \zeta_k^+\big( \xi_{k'}^-\hat q_{k'k}^{-+}\hat b_k^-\hat b_{k'}^+\tilde\rho + \xi_{k'}^{+*}\hat b_k^-\hat b_{k'}^+\tilde\rho\hat q_{k'k}^{-+} \big)\nl
    &\quad\quad - \zeta_k^-\big( \xi_{k'}^+\hat q_{k'k}^{++}\hat b_k^-\tilde\rho\hat b_{k'}^- + \xi_{k'}^{-*}\hat b_k^-\tilde\rho\hat b_{k'}^-\hat q_{k'k}^{++} \big)\nl
    &\quad\quad - \zeta_k^+\big( \xi_{k'}^-\hat q_{k'k}^{--}\hat b_{k'}^+\tilde\rho\hat b_k^+ + \xi_{k'}^{+*}\hat b_{k'}^+\tilde\rho\hat b_k^+\hat q_{k'k}^{--} \big) \Big],
\end{align}
with $\hat N_k \equiv \hat b_k^+\hat b_k^-$. By introducing the generalized Hermite conjugation, we obtain the final result, \Eq{eDQME}. 
For general conditions when the cross correlations $\la\hat F_{us}^\sigma(t)\hat F_{vs}^{\bar\sigma}(0)\ra_{\B} = \sum_{\kappa}\eta_{\kappa uvs}^\sigma e^{-\gamma_{\kappa us}^\sigma}$ exist, the extended DQME can be established by substitutions in \Eq{eDQME}:
\begin{align}
    \begin{split}
        \eta^\sigma_{us}\hat a^\sigma_{us} &\Longrightarrow \sum_v \eta^\sigma_{uvs}\hat a^\sigma_{vs}, \\
        \eta^{\bar\sigma*}_{us}\hat a^\sigma_{us} &\Longrightarrow \sum_v \eta^{\bar\sigma*}_{uvs}\hat a^\sigma_{vs}, \\
        \eta^{\sigma}_{\kappa us}\hat q^{\sigma\bar\sigma'}_{us, u's'} &\Longrightarrow \sum_v\eta^{\sigma}_{\kappa uvs}\hat q^{\sigma\bar\sigma'}_{vs, u's'}, \\
        \eta^{\bar\sigma*}_{\kappa us}\hat q^{\sigma\bar\sigma'}_{us, u's'} &\Longrightarrow \sum_v\eta^{\bar\sigma*}_{\kappa uvs}\hat q^{\sigma\bar\sigma'}_{vs, u's'}, \\
        \eta^{\sigma}_{\kappa us}\eta^{\sigma'}_{\kappa'vs'}\hat q^{\sigma\sigma'}_{us,vs'} &\Longrightarrow \sum_{u'v'}\eta^{\sigma}_{\kappa uu's}\eta^{\sigma'}_{\kappa'vv's'}\hat q^{\sigma\sigma'}_{u's,v's'}, \\
        \eta^{\bar\sigma*}_{\kappa us}\eta^{\bar\sigma'*}_{\kappa'vs'}\hat q^{\sigma\sigma'}_{us,vs'} &\Longrightarrow \sum_{u'v'}\eta^{\bar\sigma}_{\kappa uu's}\eta^{\bar\sigma'*}_{\kappa'vv's'}\hat q^{\sigma\sigma'}_{u's,v's'}.
    \end{split}
\nonumber
\end{align}
\end{widetext}

\section{Derivation of \Eq{spe222}}\label{thappA}

In this appendix, we briefly summarize the derivation of \Eq{spe222} from \Eq{spe}, cf.\,Refs.\citenum{Cas09109} and \citenum{Ana1743705} for more details.
Firstly, one may replace
\begin{align}
    \sum_{{\bm k}\in {\rm BZ}} \rightarrow \frac{\sqrt{3}a^2}{2}N\int_{\rm BZ}\!\frac{\ud^2k}{(2\pi)^2}
\end{align}
 with $\sqrt{3}a^2/2$ being the area of each cell.
 Equation (\ref{spe}) is then recast as
\begin{align}
\Gamma_s(\omega)\!=\!\frac{\sqrt{3}a^2 g_0^2}{16\pi}\!\!\int_{\rm BZ}\!\ud^2k \Big[\delta(\w^{ }\!-\!\varepsilon_{\bm k})+\delta(\w^{ }\!+\!\varepsilon_{\bm k})\Big].
\end{align}
Next, using the relation
\begin{align}
    \frac{1}{\w+i0^{+}}={\cal P}\frac{1}{\w}-i\pi \delta(\w),
\end{align}
with $\cal P$ denoting the principle part, we have 
\begin{align}
\!\!\Gamma_s(\omega)=\!-\frac{\sqrt{3}a^2 g_0^2 \w^{ }}{8\pi^2}\!\!\int_{\rm BZ}\!\ud^2k\,
{\rm Im}\frac{1}{(\w^{ }+i0^{+})^2-\varepsilon_{\bm k}^2}. 
\end{align} 
Double the domain of
integration to make it rectangular, $-2\pi/(\sqrt{3}a)\leq k_x\leq 2\pi/(\sqrt{3}a)$ and $-2\pi/a\leq k_y\leq 2\pi/a$, and change the variables $x=(\sqrt{3}a/2)k_x$ and $y=(a/2)k_y$, we obtain [cf.\,\Eq{hgra11} with \Eq{theta}]
\begin{align}\label{gmaama11}
&\Gamma_s(\omega)=-\frac{g_0^2 \w^{ }}{4\pi^2}{\rm Im}\int_{-\pi}^{\pi}\int_{-\pi}^{\pi}\!\ud x\ud y
\,\nl
&\quad\qquad\times\frac{1}{(\w^{ }+i0^{+})^2-g^2[\sin^2x+(\cos x + 2\cos y)^2]}. 
\end{align}

Since
\be 
\int_{-\pi}^{\pi}\frac{\ud x}{a-b\cos x}=\frac{2\pi}{\sqrt{a^2-b^2}},
\ee
\Eq{gmaama11} can be further simplified as
\begin{align}\label{gmaama111}
\!\!\Gamma_s(\omega)\!=\!-\frac{\w^{ }g_0^2 }{4\pi g^2}{\rm Im}\!\!\int_{-\pi}^{\pi}\!\!
 \frac{\ud y}{\sqrt{[\kappa(\w)\!-\!\cos(2y)]^2\!-\!4\cos^2 y}},
\end{align}
with $\kappa(\w) \equiv [(\w^{ }+i0^{+})^2-3g^2]/(2g^2)$.
Noting that 
\be 
\int_{-\pi}^{\pi}\ud y \,f(\cos 2y)=4\int_{0}^{\frac{\pi}{2}}\!\!\ud y \,f(\cos 2y)
\ee
for any function $f$,  we can change the variable $y$ into $z=\cos^2y$ and  recast \Eq{gmaama111} as
\begin{align}
\!\Gamma_s(\omega)\!=\!-\frac{\w^{ }g_0^2 }{4\pi g^2}{\rm Im}\!\!\int_{0}^{1}\! \frac{\ud z}{\!\sqrt{z(1-z)(z_+^{2}\!-z)(z_-^{2}\!-z)}}
\end{align}
with 
$ 
z_{\pm}(\w)=(\w^{ }+i0^{+}\pm g)/(2g)
$.
The $\Gamma_s(\omega)$ can then be  evaluated in term of an elliptic integral of the first kind as 
\begin{align}\label{gmaama12}
\Gamma_s(\omega)%=
&=-\frac{\zeta g_0^2 }{2\pi g}{\rm Im}\Bigg[ \frac{1}{\sqrt{-R(-\zeta)}} \mathbb K\left(\sqrt{\frac{-\zeta }{R(-\zeta)}}\,\right)\Bigg],
\end{align}
where $\zeta= (\w^{ }+i0^{+})/g$. Here, the functions $R(\zeta)$ and $\mathbb K(x)$ are defined in \Eqs{Rzeta} and (\ref{1stk}), respectively. 

As evident in \Eq{hgra11} with \Eq{theta}, $0\leq \w/g\leq 3$.
When $0 \leq \w/g \leq 1$ so that the argument of the elliptic function $\mathbb K$ is imaginary, we use the relation
\be 
\mathbb K(iz)=\frac{1}{\sqrt{1+z^2}}\mathbb K \left(\sqrt{\frac{z^2}{z^2+1}}\right).
\ee
For
$1 < \w/g  \leq 3$ when the argument of the elliptic function $\mathbb K$ is real, we use
\be 
\mathbb K(z)=\frac{1}{z}\left[\mathbb K \Big(\frac{1}{z}\Big)-i\mathbb K \left(\sqrt{1-\frac{1}{z^2}}\right)\right],
\ee
where $z\geq 1$ is satisfied in case $1 < \w/g  \leq 3$. 
Noting that $R(\zeta)=R(-\zeta)+\zeta$, we finally obtain \Eq{spe222} with \Eq{D}.

% \bibliographystyle{./aip}
% \bibliography{./bibrefs.bib}

\begin{thebibliography}{10}

  \bibitem{Sof12115405}
  J.~O. Sofo, G.~Usaj, P.~S. Cornaglia, A.~M. Suarez, A.~D. Hern\'andez-Nieves, and C.~A. Balseiro,
  \newblock Phys. Rev. B {\bf 85}, 115405 (2012).
  
  \bibitem{Pik14115428}
  N.~A. Pike and D.~Stroud,
  \newblock Phys. Rev. B {\bf 89}, 115428 (2014).
  
  \bibitem{Ynd14245420}
  F.~Yndurain,
  \newblock Phys. Rev. B {\bf 90}, 245420 (2014).
  
  \bibitem{Shi19125158}
  Z.~Shi, E.~M. Nica, and I.~Affleck,
  \newblock Phys. Rev. B {\bf 100}, 125158 (2019).
  
  \bibitem{Cao20045402}
  S.~Cao, C.~Cao, S.~Tian, and J.-H. Chen,
  \newblock Phys. Rev. B {\bf 102}, 045402 (2020).
  
  \bibitem{Gon16437}
  H.~{Gonz{\'a}lez-Herrero}, J.~M. {G{\'o}mez-Rodr{\'i}guez}, P.~Mallet, M.~Moaied, J.~J. Palacios, C.~Salgado, M.~M. Ugeda, J.-Y. Veuillen, F.~Yndurain, and I.~Brihuega,
  \newblock Science {\bf 352}, 437 (2016).
  
  \bibitem{Kon6437}
  J.~Kondo,
  \newblock Prog. Theor. Phys. {\bf 32}, 37 (1964).
  
  \bibitem{Kon70183}
  J.~Kondo,
  \newblock Phys. Stat. Sol. {\bf 23}, 183 (1970).
  
  \bibitem{Pus0145328}
  M.~Pustilnik and L.~I. Glazman,
  \newblock Phys. Rev. B {\bf 64}, 45328 (2001).
  
  \bibitem{Swi03195318}
  R.~Swirkowicz, J.~Barnas, and M.~Wilcznski,
  \newblock Phys. Rev. B {\bf 68}, 195318 (2003).
  
  \bibitem{Hew93}
  A.~C. Hewson,
  \newblock {\em The Kondo Problem to Heavy Fermions},
  \newblock Cambridge University Press, Cambridge, 1993.
  
  \bibitem{Mak14207401}
  K.~F. Mak, F.~H. da~Jornada, K.~He, J.~Deslippe, N.~Petrone, J.~Hone, J.~Shan, S.~G. Louie, and T.~F. Heinz,
  \newblock Phys. Rev. Lett. {\bf 112}, 207401 (2014).
  
  \bibitem{Soh14125414}
  T.~Sohier, M.~Calandra, C.-H. Park, N.~Bonini, N.~Marzari, and F.~Mauri,
  \newblock Phys. Rev. B {\bf 90}, 125414 (2014).
  
  \bibitem{Cre15635019}
  N.~Creange, C.~Constantin, J.-X. Zhu, A.~V. Balatsky, and J.~T. Haraldsen,
  \newblock Advances in Condensed Matter Physics {\bf 2015}, 635019 (2015).
  
  \bibitem{Bul16270}
  L.~Bulusheva, O.~Sedelnikova, and A.~Okotrub,
  \newblock Int. J. Quantum Chem. {\bf 116}, 270 (2016).
  
  \bibitem{Anv21155402}
  R.~Anvari, E.~Zaremba, and M.~M. Dignam,
  \newblock Phys. Rev. B {\bf 104}, 155402 (2021).
  
  \bibitem{Wil75773}
  K.~G. Wilson,
  \newblock Rev. Mod. Phys. {\bf 47}, 773 (1975).
  
  \bibitem{Yos909403}
  M.~Yoshida, M.~A. Whitaker, and L.~N. Oliveira,
  \newblock Phys. Rev. B {\bf 41}, 9403 (1990).
  
  \bibitem{Cos973003}
  T.~A. Costi,
  \newblock Phys. Rev. B {\bf 55}, 3003 (1997).
  
  \bibitem{Bul988365}
  R.~Bulla, A.~C. Hewson, and T.~Pruschke,
  \newblock J. Phys.: Cond. Matt. {\bf 10}, 8365 (1998).
  
  \bibitem{Pet06245114}
  R.~Peters, T.~Pruschke, and F.~B. Anders,
  \newblock Phys. Rev. B {\bf 74}, 245114 (2006).
  
  \bibitem{And08195216}
  F.~B. Anders,
  \newblock J. Phys.: Condens. Matter {\bf 20}, 195216 (2008).
  
  \bibitem{Gul11349}
  E.~Gull, A.~J. Millis, A.~I. Lichtenstein, A.~N. Rubtsov, M.~Troyer, and P.~Werner,
  \newblock Rev. Mod. Phys. {\bf 83}, 349 (2011).
  
  \bibitem{Har15085430}
  R.~H\"artle, G.~Cohen, D.~R. Reichman, and A.~J. Millis,
  \newblock Phys. Rev. B {\bf 92}, 085430 (2015).
  
  \bibitem{Mey9073}
  H.~D. Meyer, U.~Manthe, and L.~Cederbaum,
  \newblock Chem. Phys. Lett. {\bf 165}, 73  (1990).
  
  \bibitem{Wan031289}
  H.~B. Wang and M.~Thoss,
  \newblock J. Chem. Phys. {\bf 119}, 1289 (2003).
  
  \bibitem{Zhe25052051}
  J.~Zheng, Y.~Xie, J.~Peng, Z.~Han, and Z.~Lan,
  \newblock J. Chem. Phys. {\bf 162}, 052501 (2025).
  
  \bibitem{And05196801}
  F.~B. Anders and A.~Schiller,
  \newblock Phys. Rev. Lett. {\bf 95}, 196801 (2005).
  
  \bibitem{Fri06144410}
  L.~Fritz, S.~Florens, and M.~Vojta,
  \newblock Phys. Rev. B {\bf 74}, 144410 (2006).
  
  \bibitem{Nis04613}
  S.~Nishimoto and E.~Jeckelmann,
  \newblock J. Phys.: Condens. Matter {\bf 16}, 613 (2004).
  
  \bibitem{Mer12075153}
  L.~Merker, A.~Weichselbaum, and T.~A. Costi,
  \newblock Phys. Rev. B {\bf 86}, 075153 (2012).
  
  \bibitem{Ren22e1614}
  J.~Ren, W.~Li, T.~Jiang, Y.~Wang, and Z.~Shuai,
  \newblock WIREs Comput. Mol. Sci. , e1614 (2022).
  
  \bibitem{Pri10050404}
  J.~Prior, A.~W. Chin, S.~F. Huelga, and M.~B. Plenio,
  \newblock Phys. Rev. Lett. {\bf 105}, 050404 (2010).
  
  \bibitem{Nus20155134}
  A.~N\"u\ss{}eler, I.~Dhand, S.~F. Huelga, and M.~B. Plenio,
  \newblock Phys. Rev. B {\bf 101}, 155134 (2020).
  
  \bibitem{Fey63118}
  R.~P. Feynman and F.~L. \mbox{Vernon, Jr.},
  \newblock Ann. Phys. {\bf 24}, 118 (1963).
  
  \bibitem{Mak92435}
  N.~Makri,
  \newblock Chem. Phys. Lett. {\bf 193}, 435 (1992).
  
  \bibitem{Mak954600}
  N.~Makri and D.~E. Makarov,
  \newblock J. Chem. Phys. {\bf 102}, 4600 (1995).
  
  \bibitem{Mak954611}
  N.~Makri and D.~E. Makarov,
  \newblock J. Chem. Phys. {\bf 102}, 4611 (1995).
  
  \bibitem{Jor19240602}
  M.~R. J\o{}rgensen and F.~A. Pollock,
  \newblock Phys. Rev. Lett. {\bf 123}, 240602 (2019).
  
  \bibitem{Ric22167403}
  M.~Richter and S.~Hughes,
  \newblock Phys. Rev. Lett. {\bf 128}, 167403 (2022).
  
  \bibitem{Cyg22662}
  M.~Cygorek, M.~Cosacchi, A.~Vagov, V.~M. Axt, B.~W. Lovett, J.~Keeling, and E.~M. Gauger,
  \newblock Nat. Phys. {\bf 18}, 662 (2022).
  
  \bibitem{Coh15266802}
  G.~Cohen, E.~Gull, D.~R. Reichman, and A.~J. Millis,
  \newblock Phys. Rev. Lett. {\bf 115}, 266802 (2015).
  
  \bibitem{Che17054105}
  H.-T. Chen, G.~Cohen, and D.~R. Reichman,
  \newblock The Journal of chemical physics {\bf 146}, 054105 (2017).
  
  \bibitem{Cai202430}
  Z.~Cai, J.~Lu, and S.~Yang,
  \newblock Communications on Pure and Applied Mathematics {\bf 73}, 2430 (2020).
  
  \bibitem{Ber21155104}
  C.~Bertrand, D.~Bauernfeind, P.~T. Dumitrescu, M.~Ma\ifmmode~\check{c}\else \v{c}\fi{}ek, X.~Waintal, and O.~Parcollet,
  \newblock Phys. Rev. B {\bf 103}, 155104 (2021).
  
  \bibitem{Sor19043303}
  M.~E. Sorantin, D.~M. Fugger, A.~Dorda, W.~von~der Linden, and E.~Arrigoni,
  \newblock Phys. Rev. E {\bf 99}, 043303 (2019).
  
  \bibitem{Yan14054105}
  Y.~J. Yan,
  \newblock J. Chem. Phys. {\bf 140}, 054105 (2014).
  
  \bibitem{Zha15024112}
  H.~D. Zhang, R.~X. Xu, X.~Zheng, and Y.~J. Yan,
  \newblock J. Chem. Phys. {\bf 142}, 024112 (2015).
  
  \bibitem{Wan22170901}
  Y.~Wang and Y.~J. Yan,
  \newblock J. Chem. Phys. {\bf 157}, 170901 (2022).
  
  \bibitem{Tan906676}
  Y.~Tanimura,
  \newblock Phys. Rev. A {\bf 41}, 6676 (1990).
  
  \bibitem{Yan04216}
  Y.~A. Yan, F.~Yang, Y.~Liu, and J.~S. Shao,
  \newblock Chem. Phys. Lett. {\bf 395}, 216 (2004).
  
  \bibitem{Xu05041103}
  R.~X. Xu, P.~Cui, X.~Q. Li, Y.~Mo, and Y.~J. Yan,
  \newblock J. Chem. Phys. {\bf 122}, 041103 (2005).
  
  \bibitem{Jin08234703}
  J.~S. Jin, X.~Zheng, and Y.~J. Yan,
  \newblock J. Chem. Phys. {\bf 128}, 234703 (2008).
  
  \bibitem{Tan20020901}
  Y.~Tanimura,
  \newblock J. Chem. Phys. {\bf 153}, 020901 (2020).
  
  \bibitem{Su23024113}
  Y.~Su, Z.-H. Chen, Y.~Wang, X.~Zheng, R.-X. Xu, and Y.~Yan,
  \newblock J. Chem. Phys. {\bf 159}, 024113 (2023).
  
  \bibitem{Phi12}
  P.~Phillips,
  \newblock {\em Advanced Solid State Physics},
  \newblock Cambridge University Press, 2012.
  
  \bibitem{Bal06373}
  A.~V. Balatsky, I.~Vekhter, and J.-X. Zhu,
  \newblock Rev. Mod. Phys. {\bf 78}, 373 (2006).
  
  \bibitem{Moc21186804}
  C.~P. Moca, I.~Weymann, M.~A. Werner, and G.~Zar\'and,
  \newblock Phys. Rev. Lett. {\bf 127}, 186804 (2021).
  
  \bibitem{Li24032620}
  X.~Li, S.-X. Lyu, Y.~Wang, R.-X. Xu, X.~Zheng, and Y.~J. Yan,
  \newblock Phys. Rev. A {\bf 110}, 032620 (2024).
  
  \bibitem{Che22221102}
  Z.~H. Chen, Y.~Wang, X.~Zheng, R.~X. Xu, and Y.~J. Yan,
  \newblock J. Chem. Phys. {\bf 156}, 221102 (2022).
  
  \bibitem{Liu17024110}
  J.~Liu,
  \newblock J. Chem. Phys. {\bf 146}, 024110 (2017).
  
  \bibitem{Xu18114103}
  R.~X. Xu, Y.~Liu, H.~D. Zhang, and Y.~J. Yan,
  \newblock J. Chem. Phys. {\bf 148}, 114103 (2018).
  
  \bibitem{Che23074102}
  Z.-H. Chen, Y.~Wang, R.-X. Xu, and Y.~Yan,
  \newblock J. Chem. Phys. {\bf 158}, 074102 (2023).
  
  \bibitem{Got152722}
  F.~Gottwald, S.~D. Ivanov, and O.~Kühn,
  \newblock J. Phys. Chem. Lett. {\bf 6}, 2722 (2015).
  
  \bibitem{Yao212130010}
  X.~Yao,
  \newblock Int. J. Mod. Phys. A {\bf 36}, 2130010 (2021).
  
  \bibitem{Wei25_arxiv_2501_05562}
  Z.-Y. Wei, T.~Shi, J.~I. Cirac, and E.~A. Demler,
  \newblock arXiv: 2501.05562  (2025).
  
  \bibitem{Yan19074106}
  Y.~A. Yan,
  \newblock J. Chem. Phys. {\bf 150}, 074106 (2019).
  
  \bibitem{Liu17102319}
  X.~Liu and J.~Liu,
  \newblock J. Chem. Phys. {\bf 148}, 102319 (2017).
  
  \bibitem{He22e1619}
  X.~He, B.~Wu, Y.~Shang, B.~Li, X.~Cheng, and J.~Liu,
  \newblock WIREs Comput. Mol. Sci. {\bf 12}, e1619 (2022).
  
  \bibitem{Wu24644}
  B.~Wu, X.~He, and J.~Liu,
  \newblock J. Phys. Chem. Lett. {\bf 15}, 644 (2024).
  
  \bibitem{He245452}
  X.~He, X.~Cheng, B.~Wu, and J.~Liu,
  \newblock J. Phys. Chem. Lett. {\bf 15}, 5452 (2024).
  
  \bibitem{Cas09109}
  A.~H. {Castro Neto}, N.~M.~R. Peres, K.~S. Novoselov, and A.~K. Geim,
  \newblock Rev. Mod. Phys. {\bf 81}, 109 (2009).
  
  \bibitem{Ana1743705}
  V.~Ananyev and M.~Ovchynnikov,
  \newblock Condens. Matter Phys. {\bf 20} (2017).
  
  \bibitem{Ye16608}
  L.~Z. Ye, X.~L. Wang, D.~Hou, R.~X. Xu, X.~Zheng, and Y.~J. Yan,
  \newblock WIREs Comput. Mol. Sci. {\bf 6}, 608 (2016).
  
  \end{thebibliography}

\end{document}